\documentclass{bmcart}

\usepackage[utf8]{inputenc} 

\usepackage[sort,compress]{natbib}
\usepackage{amsmath,amssymb,bm}
\usepackage{graphicx}
\usepackage{hyperref}

\startlocaldefs

\newtheorem{mydef}{Definition}
\newtheorem{mythe}{Proposition}

\DeclareMathOperator*{\polylog}{polylog}

\endlocaldefs

\begin{document}

\begin{frontmatter}

\begin{fmbox}
\dochead{Research}


\title{Applying compressed sensing to genome-wide association studies}


\author[
   addressref={aff1},                   
   email={vattikutis@niddk.nih.gov}   
]{\inits{SV}\fnm{Shashaank} \snm{Vattikuti}}
\author[
   addressref={aff1,aff2,aff5},
   email={leex2293@umn.edu}
]{\inits{JJL}\fnm{James J} \snm{Lee}}
\author[
   addressref={aff3,aff5},
   email={chrchang@alumni.caltech.edu}
]{\inits{Christopher CC}\fnm{Christopher C} \snm{Chang}}
\author[
   addressref={aff4,aff5},
   noteref={n1},   
   email={hsu@msu.edu}
]{\inits{SDHH}\fnm{Stephen D H} \snm{Hsu}}
\author[
   addressref={aff1},
   corref={aff1},
   noteref={n1},   
   email={carsonc@mail.nih.gov}
]{\inits{Carson CC}\fnm{Carson C} \snm{Chow}}


\address[id=aff1]{
  \orgname{Laboratory of Biological Modeling, National Institute of Diabetes and Digestive and Kidney Diseases, National Institutes of Health}, 
   \street{12 Center Drive},                     %
  \postcode{20814}                                
  \city{Bethesda, MD},                              
  \cny{USA}                                    
}

\address[id=aff2]{%
  \orgname{Department of Psychology, University of Minnesota Twin Cities},
  \street{75 East River Parkway},
  \postcode{55455}
  \city{Minneapolis, MN},
  \cny{USA}
}

\address[id=aff3]{%
  \orgname{BGI Hong Kong},
  \street{16 Dai Fu Street, Tai Po Industrial Estate},
  \city{Hong Kong}
}

\address[id=aff4]{%
  \orgname{Office of the Vice President for Research and Graduate Studies, Michigan State University},
  \street{426 Auditorium Road},
  \postcode{48824}
  \city{East Lansing, MI},
  \cny{USA}
}

\address[id=aff5]{%
  \orgname{Cognitive Genomics Lab, BGI Shenzhen},
  \street{Yantian District},
  \city{Shenzhen},
  \cny{China}
}


\begin{artnotes}
\note[id=n1]{Correspondences to hsu@msu.edu or carsonc@mail.nih.gov} 
\end{artnotes}

\end{fmbox}


\begin{abstractbox}

\begin{abstract} 
\parttitle{Background}
The aim of a genome-wide association study (GWAS) is to isolate DNA
markers for variants affecting phenotypes of interest. Linear
regression is employed for this purpose, and in recent years a
signal-processing paradigm known as compressed sensing (CS) has
coalesced around a particular class of regression techniques. CS is
not a method in its own right, but rather a body of theory regarding
signal recovery when the number of predictor variables (i.e., genotyped markers) exceeds the sample size. 
\parttitle{Results}
Using CS theory, we show that all markers with nonzero coefficients
can be identified (selected) using an efficient algorithm, provided
that they are sufficiently few in number (sparse) relative to sample
size.  For heritability $h^2 = 1$, there is a sharp phase transition from poor performance 
to complete selection as the sample size is increased. For
heritability values less than one, complete selection can still occur,
although the transition is smoothed. The transition boundary is only weakly dependent on the
total number of genotyped markers. In the presence of correlations among predictor variables
(linkage disequilibrium), measures of recovery such as the squared
deviations of the estimated coefficients from their true values are
also smoothed. More practical measures of signal recovery can
accommodate linkage disequilibrium between a true causal variant and markers residing in the same genomic region,
and indeed such measures (e.g., median $P$-value of selected markers)
continue to show the good behavior expected in the absence of linkage
disequilibrium. When applying this approach to the
  GWAS analysis of height, we show that 70-100\% of the selected
  markers are strongly correlated with height-associated markers identified by the GIANT Consortium.

\parttitle{Conclusion}
The signal-processing paradigm known as CS is applicable
to GWAS. The crossing of a transition
boundary between distinct phases provides an objective means of determining when true
trait-associated markers are being recovered, and we propose a novel analysis strategy that takes advantage of this property. The median $P$-value exhibits a sharp transition as the sample size is increased, indicating nearly complete recovery of true signal (causal variants or nearby proxy markers).  In addition, given a limited sample size, it may still be possible to discover a phase
transition by increasing the penalization, although in this case only a subset of the support can be recovered. Supposing that the recovery of the entire set is desired, we find for $h^2 \sim 0.5$ that a sample size of approximately thirty times the number of markers with nonzero coefficients is sufficient.

\end{abstract}




\end{abstractbox}
%

\end{frontmatter}




\section*{Background}

The search for genetic variants associated with a given phenotype in a
genome-wide association study (GWAS) is a classic example of what has
been called a $p \gg n$ problem, where $n$ is the sample size (number
of subjects) and $p$
is the number of predictor variables (genotyped markers) \cite{johnstone:2009}. Estimating the partial regression coefficients of the predictor variables by ordinary least squares (OLS)  requires that the sample size exceed the number
of coefficients, which in the GWAS context may be of order $10^5$ or even $10^6$. 
The difficulty of assembling such large samples has been one obstacle hindering the simultaneous estimation of all regression coefficients advocated by some authors \cite{hoggart:2008,goddard:2009:shrink,kemper:2012:sheep}.

The typical procedure in GWAS is to estimate each coefficient by OLS
independently and retain those meeting a strict threshold; this approach is sometimes called \emph{marginal regression} (MR) \cite{genovese:2012}. Although the implementation of MR in GWAS has led to an avalanche of
discoveries \cite{visscher:2012:gwas}, it is uncertain whether it will
be optimal as datasets continue to increase in size. Many genetic markers associated with a trait are likely to be missed because they do not pass the chosen significance threshold \cite{yang:2010:snpherit}.

Unlike MR, which directly estimates whether each coefficient is nonzero, an $L_1$-penalization algorithm such as the lasso effectively translates the estimates toward the origin, where many are truncated out of the model \cite{tibshirani:1996}. If the number of variants associated with a typical complex trait is indeed far fewer than the total number of polymorphic sites \cite{park:2011,stahl:2012,ripke:2013}, then it is reasonable to believe that $L_1$ penalization will at least be competitive with MR. Methods relying on the assumption of \emph{sparsity} (few nonzero coefficients relative to sample size) have in fact been
adopted by workers in the field of genomic selection (GS), which uses genetic information
to guide the artificial selection of livestock and crops
\cite{meuwissen:2001,deloscampos:2010,hayes:2010,meuwissen:2013}. Note that the aim of GS (phenotypic prediction) is somewhat distinct from that of GWAS (the identification of markers tagging causal variants). The lasso is one of the methods studied by GS investigators \cite{usai:2009,wimmer:2013}, although Bayesian methods that regularize the coefficients with strong priors tend to be favored \cite{zhou:2013:bslmm,gianola:2013:gs}.

In this paper we show that  theoretical
results from the field of \emph{compressed sensing} (CS) supply a
rigorous quantitative framework for the application of regularization
methods to GWAS. In particular, CS theory provides a mathematical justification for the use of $L_1$-penalized regression to recover sparse vectors of coefficients and highlights the difference between \emph{selection} of the markers with nonzero coefficients and the \emph{fitting} of the precise coefficient values. CS theory also addresses the robustness of $L_1$ algorithms to the distribution of nonzero coefficient magnitudes.

Besides supplying a rule of thumb for the sample size
sufficing to select the markers with true nonzero coefficients, CS
gives an independent quantitative criterion for determining whether a
given dataset has in fact attained that sample size. That is, whereas
biological assumptions regarding the number of nonzeros do enter into
the rule of thumb about sample size, these assumptions need not hold
for the use of $L_1$ penalization to be justified; this is because the
returned results themselves inform the investigator whether the
assumptions are met. 

We emphasize that CS is not a method \emph{per se} but may be considered a general theory of regression
  that takes into account model complexity (sparsity). The theory is
  still valid in the classical regression domain of $n>p$ but establishes
  conditions for when full recovery of nonzero coefficients is still possible when $n<p$ \cite{donoho:2005,candes:2009,candes:2011:ripless}.
   Our work
therefore should not be directly compared to recent literature
proposing and evaluating GS methods
\cite{zhou:2013:bslmm,gianola:2013:gs}. Rather, our goal is  to elucidate
properties of well-known methods, already in use by GWAS and GS
researchers, whose mathematical attributes and empirical prospects may
be insufficiently appreciated. 

Using more than 12,000 subjects from
the ARIC European American and GENEVA cohorts
and nearly 700,000 single-nucleotide polymorphisms (SNPs), we show that the matrix of genotypes acquired
in GWAS obeys properties suitable for the application of CS
theory. In particular, a given sample size determines the maximum
number of nonzeros that will be fully selected using an $L_1$-penalization regression algorithm. If
the sample size is too small, then the complete set of
nonzeros will not be selected.
The transition between poor and complete selection is sharp
in the noiseless case (narrow-sense heritability equal to one). It is smoothed in the presence of
noise (heritability less than one) but still fully detectable. Consistent with CS theory, we find in cases with realistic residual noise that the minimal sample
size for full recovery is primarily determined by the number of nonzeros, depends
very weakly on the number of genotyped markers 
\cite{candes:2006:rob,donoho:2011,candes:2011:ripless}, and is robust
to the distribution of coefficient magnitudes
\cite{donoho:2009}. 

\subsection*{Theory of Compressed Sensing}

The linear model of quantitative genetics is
\begin{equation}\label{model}
\mathbf{y} = \mathbf{Ax}+\mathbf{e},
\end{equation}
where $\mathbf{y}\in\mathbb{R}^{n}$ is the vector of
phenotypes, $\mathbf{A}\in\mathbb{R}^{n\textrm{x}p}$ is the matrix of
standardized genotypes,
$\mathbf{x}\in\mathbb{R}^{p}$ is the vector of partial regression coefficients, and $\mathbf{e}\in\mathbb{R}^{n\textrm{x}1}$ is the vector of residuals. In the CS literature, $\mathbf{A}$ is often called the \emph{sensing} or \emph{measurement} matrix. Standardizing $\mathbf{A}$ does not affect the results and makes it simpler to utilize CS theory. We suppose that $\mathbf{x}$ contains $s$ nonzero coefficients (``nonzeros'') whose indices we wish to know.

The phase transition to complete selection is best quantified with two ratios $(\rho, \delta)$, where  $\rho=s/n$ is a measure of the sparsity of nonzeros with respect to the sample size and $\delta=n/p$ is a measure of the undersampling. If we plot $\delta$ on the abscissa ($x$-axis) and $\rho$ on the ordinate ($y$-axis), we have a \emph{phase plane} on the square $(0, 1) \times (0, 1)$, where each point represents a possible GWAS situation (sample size, number of genotyped markers, number of true nonzeros). The performance of any given method can be assessed by evaluating a measure of recovery quality at each point of the plane.  For an arbitrary $p$-vector $\mathbf{x}$, we use the following notation for the $L_1$ and $L_2$ norms:
\begin{equation*}
\| \mathbf{x} \|_{L_1} = \sum_{i=1}^p | x_i | \quad \textrm{and} \quad \| \mathbf{x} \|_{L_2} = \sqrt{ \sum_{i=1}^p x_i^2 }. 
\end{equation*}

Our results rely on two lines of research in the field of CS, which we
summarize as two propositions.\\

\begin{mythe}\label{donoho_theorem}~\cite{donoho:2005,donoho:2006:neighbor,donoho:2009:ptc_igor,donoho:2011}
Suppose that the entries of the sensing matrix $\mathbf{A}$ are drawn from independent normal distributions and $\mathbf{e}$ is the zero vector (noiseless
case). Then the $\rho -\delta$ plane is partitioned by a curve $\rho =
\rho_{L_1}(\delta)$ into two phases. Below the curve the solution of
$\min\limits_{\hat{\mathbf{x}}} \| \hat{\mathbf{x}} \|_{L_1}$ subject to $\mathbf{A}\hat{\mathbf{x}} = \mathbf{y}$ leads to $\hat{\mathbf{x}} = \mathbf{x}$ with probability converging to one as $n, p, s \to \infty$ in such a way that $\rho$ and $\delta$ remain constant. Above the curve $\hat{\mathbf{x}} \neq \mathbf{x}$ with similarly high probability.
\end{mythe}
\phantom{1}
The function $\rho_{L_1}(\delta)$ can be analytically calculated \cite{donoho:2006:neighbor}. 
Although Figure~\ref{fig:rhodelta_random}A presents some of our
empirical results, which we will discuss below, it can be
taken as an illustration of the meaning of Proposition~\ref{donoho_theorem}. The color scale represents the goodness of recovery, and
the black curve is the graph of $\rho_{L_1}(\delta)$. It can be seen that increasing the sample size relative to $s$ (decreasing $\rho$) leads to a sharp transition from poor to good recovery for $\delta<<1$ (i.e. $n << p$). In other words, despite the fact that solving for $\mathbf{x}$ in $\mathbf{Ax} = \mathbf{y}$ is strictly speaking underdetermined given $n < p$, minimizing $|| \hat{\mathbf{x}} ||_{L_1}$ subject to the system of equations still yields recovery of $\mathbf{x}$ with high probability if $n$ is sufficiently large relative to $s$.

Most phenotypes do not have a heritability of one and thus are not noiseless, but CS theory shows that selection is still possible in this situation.  Before stating the relevant CS result, we need to define two quantities characterizing the genotype matrix $\mathbf{A}$.\\

\begin{mydef}~\cite{candes:2011:ripless}
The matrix $\mathbf{A}$ satisfies \emph{isotropy} if the expectation value of $\mathbf{A}^\prime \mathbf{A}$ is equal to the identity matrix.
\end{mydef}
\phantom{1}
In the context of GWAS, a matrix of gene counts is isotropic if all markers are in linkage equilibrium (LE). \\

\begin{mydef}~\cite{candes:2011:ripless}
The \emph{coherence} of the matrix $\mathbf{A}$ is the smallest number $\gamma$ such that, for each row $\mathbf{a}$ of the matrix, 
\begin{equation*}
\max_{1 \leq t \leq p} |\mathbf{a}_t|^2 \leq \gamma. 
\end{equation*}
\end{mydef}
A matrix of genotypes is thus reasonably \emph{incoherent} if
the magnitudes of the matrix elements do not differ greatly from each
other. In the GWAS context, $\mathbf{A}$ will be reasonably incoherent
if all markers with very low minor allele frequency (MAF) are
pruned, since $\mathbf{A}$ is standardized and the standard deviation scales with MAF.

We can now state \\

\begin{mythe}\label{candes_theorem}~\cite{candes:2011:ripless}
Suppose that the sensing matrix $\mathbf{A}$ is isotropic with coherence $\gamma$. If $n > C \, \gamma \, s \log p$ for a constant $C$ then the solution of the problem
\begin{equation*}
\min_{\hat{\mathbf{x}}} \bigg[ \| \mathbf{y} - \mathbf{A}\hat{\mathbf{x}} \|^2_{L_2} + \lambda \| \hat{\mathbf{x}} \|_{L_1} \bigg]
\end{equation*}
with a suitable choice of $\lambda$ obeys
\begin{equation*}
\| \hat{\mathbf{x}} - \mathbf{x} \|^2_{L_2} \leq  \frac{\sigma^2_E}{n} s \polylog p,
\end{equation*}
where $\sigma^2_E$ is the variance of the residuals in $\mathbf{e}$. 
\end{mythe}
\phantom{1}
Two features of Proposition~\ref{candes_theorem} are worth
noting. First, no strong restrictions on $\mathbf{x}$ are required. Second, the critical threshold value of $n$ depends linearly on
$s$ but only (poly)logarithmically on $p$.  For $n$ larger than the critical
value, the deviations of the estimated coefficients from the true values will follow the expected OLS scaling of $1/\sqrt{n}$.

These results are more powerful than they might seem from the
restrictive hypotheses required for brief formulations. For example,
it has been shown that a curve similar to that in
Proposition~\ref{donoho_theorem} also demarcates a phase transition in
the case of $\mathbf{e} \neq \mathbf{0}$ --- although, as might be
expected from a comparison of Propositions~\ref{donoho_theorem}
and~\ref{candes_theorem}, with large residual noise the transition is
to a regime of gradual improvement with $n$ rather than to
instantaneous recovery \cite{donoho:2006:break,donoho:2011}. A remarkable feature of this gradual improvement, however, should be noted. Proposition~\ref{candes_theorem} states that the scaling of the total fitting error in the favorable regime is within a (poly)logarithmic factor of what would have been achieved if the identities of the $s$ nonzeros had been revealed in advance by an oracle. This result implies that perfect selection of nonzeros can occur before the magnitudes of the coefficients are well fit.
Even if the residual noise is substantial enough to prevent the sharp transition from large to negligible fitting error evident in Figure~\ref{fig:rhodelta_random}A, the total magnitude of the error in the favorable phase is little larger than what would be expected given perfect selection of the nonzeros.

 Recent
work has also generalized the sensing matrix,
  $\mathbf{A}$, in Proposition~\ref{donoho_theorem} to several non-normal distributions (although not to genotype matrices {\it per se}) \cite{donoho:2009:ptc_igor,monajemi:2013}. Furthermore, the form of Proposition~\ref{candes_theorem} also holds under a weaker form of isotropy that allows the expectation of $\mathbf{A}^\prime \mathbf{A}$ to differ from the identity matrix by a small quantity (see \cite{candes:2011:ripless} for the specification of the matrix norm). The latter generalization is promising because the covariance matrix in GWAS deviates toward block-diagonality as a result of linkage disequilibrium (LD) among spatially proximate variants.
  
Whereas the penalization parameter
$\lambda$ in Proposition~\ref{candes_theorem} is often determined
empirically through cross-validation, CS places a theoretical lower
bound on its value that is based on the magnitude of the noise
\cite{candes:2011:ripless} (referred here as $\lambda_{min}$ or $\lambda$). A special feature of the GWAS context is
that an estimate of the residual variance can be obtained from the
genomic-relatedness method \cite{yang:2010:snpherit,vattikuti:2012,vattikuti:2014,lee:2014},
thereby enabling the substitution of a theoretical noise-dependent
bound for empirical cross-validation. Such noise-dependent bounds
appear in other selection theories, including MR, and thus are not
specific to CS \cite{genovese:2012,johnstone:1998}. As noted by~\cite{johnstone:1998}, such bounds tend to be conservative. Here, we
show that the CS noise-dependent bound demonstrates good selection
properties. A data-specific method such as cross-validation may
exhibit slightly better properties but is computationally more
expensive.

Given this body of CS theory, a number of questions regarding the use of $L_1$-penalized regression in GWAS naturally arise:
\begin{enumerate}
\item Does the matrix of genotypes $\mathbf{A}$ in the GWAS setting fall into the class of matrices exhibiting the CS phase transition across the curve $\rho_{L_1}(\delta)$, as described by Proposition~\ref{donoho_theorem}? 

\item Since large residual noise is typical, we must also ask: is $\mathbf{A}$ sufficiently isotropic and incoherent to make the regime of good performance described by Proposition~\ref{candes_theorem} practically attainable? Since $\log p$ slowly varies over the relevant range of $p$, we can absorb $\gamma$ and $\log p$ into the constant factor and phrase the question more provocatively: given that $n > C s$ is required for good recovery, what is $C$?

\item In practice a measure of recovery relying on the unknown $\mathbf{x}$, such as a function of $\| \hat{\mathbf{x}} - \mathbf{x} \|_{L_2}$, cannot be used. Is there a measure of recovery, then, that depends solely on observables? 

\end{enumerate}

The aim of the present work is to answer these three questions.

\section*{Data Description}

All participants gave informed consent. All studies were approved by their appropriate Research Ethics Committees.

We used the Atherosclerosis
Risk in Community (ARIC) and Gene Environment Association Studies
(GENEVA) European American cohort. The datasets were obtained from dbGaP
at \url{http://www.ncbi.nlm.nih.gov/sites/entrez?Db=gap} through dbGaP
accession numbers [ARIC:phs000090] and [GENEVA:phs000091]. The ARIC population
consists of a large sample of unrelated individuals and some families. The population was recruited in 1987 from four
centers across the United States: Forsyth County, North Carolina;
Jackson, Mississippi; Minneapolis, Minnesota; and Washington County,
Maryland.

The ARIC subjects were genotyped with the Affymetrix Human SNP Array 6.0. We
selected biallelic autosomal markers based on a Hardy-Weinberg
equilibrium tolerance of $P < 10^ {-3} $. Preprocessing was performed with
PLINK 2 (\url{https://www.cog-genomics.org/plink2/}) \cite{purcell:2007}. 

The datasets were merged to create a SNP genotype matrix ($\mathbf{A}$) consisting of 12,464 subjects and 
693,385 SNPs. SNPs were coded by their minor allele, resulting in values of 0, 1, or 2. Each column of $\mathbf{A}$ was standardized to have zero  mean and unit variance. Missing genotypes were replaced with the mean (i.e., zero) after
standardization. We compared results for the phase
  transition for a limited number of cases when the missing genotypes
  were imputed based on sampling from a Binomial distribution and the
  respective minor allele frequency. We found no difference between
  the imputation methods for our data sets.  

We simulated phenotypes according to Equation~\ref{model}, rescaling
each term to leave the phenotypic variance equal to unity and the
variance of the breeding values in $\mathbf{Ax}$ to match the target narrow-sense
heritability $h^2$, which is the proportion of
  phenotypic variance due to additive genetic factors. For standardized phenotypes,  $h^2$ is equivalent to the additive genetic
  variance, which is defined to equal one in the noiseless case. We
  chose $h^2= 0.5$ to represent the noisy case because many human
  traits show a SNP-based heritability close to this value
  \cite{yang:2010:snpherit,davies:2011,vattikuti:2012}. 

The magnitudes of the $s$ nonzeros in $\mathbf{x}$ were
drawn from either the set $\{-1,1\}$ or hyperexponential
distributions. We defined two hyperexponential distributions
(Hyperexponential 1 and 2) and each was generated by summing two
exponentials with the same amplitude but different decay constants. The
pair of decay constants for Hyperexponential 1 were $0.05 s$
and $p$, and that of Hyperexponential 2 were $0.2 s$
and $p$. The coefficients were then truncated to keep only the top $s$ nonzero
coefficients, the rest were made zero, and 50\% of the nonzeros had negative signs. The hyperexponential form was motivated by
\cite{chatterjee:2013} but the decay constants were arbitrarily
chosen. For all coefficient ensembles, the nonzeros were randomly
distributed among the SNPs. When examining the dependence of an
outcome on $n$, $p$, and $s$ the set $p$ was either chosen randomly
across the genome without replacement or restricted to all chromosome 22 SNPs, and $n$ and
$s$ were randomly sampled without replacement. A single set of SNPs
was used for all analyses of the genomic random $p$ set.

We also considered a real phenotype (height) rather than a
simulated one, using 12,454 subjects with measurements
of height adjusted for sex. We examined different values of $n$ and
fixed $p$ by always using all markers in our dataset. A called nonzero was counted as a true positive in the numerator of our ``adjusted positive predictive value'' (to be defined later) if the marker was a member of a proxy set based on
height-associated SNPs discovered by the GIANT Consortium \cite{langoallen:2010}. The set was generated using
the BROAD SNAP database
(\url{http://www.broadinstitute.org/mpg/snap/})
\cite{johnson:2008:snap}. We based our proxy criterion on bp distance
rather than LD, as we found the correlations between SNPs in our
dataset to be larger in magnitude than those recorded in the SNAP
database. We generated a proxy list based on a maximum basepair
distance of 500 kb, which was the maximum distance that could be
queried.

\section*{Analysis}

\subsection*{Phase transition to complete selection}

We first studied the case of independent markers to gain insight into the more realistic case of LD among spatially proximate markers \cite{abraham:2013,wimmer:2013}.
In the
noiseless case ($\mathbf{e}=\mathbf{0}$), it has been proven that there is a universal phase
transition boundary between poor and complete selection in
the $\rho-\delta$ plane (Proposition~\ref{donoho_theorem}) \cite{donoho:2005,donoho:2006:neighbor,donoho:2009:ptc_igor,donoho:2011}. The existence of this boundary is largely
independent of the explicit values of $s$, $n$, and $p$ for a large class of
sensing matrices, including sensing matrices generated by the multivariate normal
distribution. The transition boundary does depend, however,
on certain properties of the distribution describing the coefficients. For example, the boundary can depend critically on whether the coefficients are all positive or can have either sign, although the particular form of the distribution within either of these two broad classes is less important. Genetic
applications typically have real-valued coefficients, which are
in the same class (i.e., in terms of phase transition properties) as coefficients drawn from the set $\{ -1, 1\}$
\cite{donoho:2009,donoho:2010:precise}, which we used in the majority of our
simulations. We also studied selection performance when the
coefficients are hyperexponentially distributed (see Data Description).

The phase
transition can be explored using multiple measures of selection
quality. Figure~\ref{fig:rhodelta_random}A shows the normalized error
($NE$) (Equation~\ref{ne}) of the coefficient estimates returned by
the $L_1$-penalized regression algorithm in our study of a simulated
phenotype and a random selection of SNPs ascertained in a real GWAS
for the noiseless case. The
boundary between poor and good performance, as evidenced by this measure, was
well approximated by the theoretically derived curve
\cite{donoho:2006:neighbor}, confirming that a matrix of independent
SNPs ascertained in GWAS qualifies as a CS sensing matrix. 

The noiseless case corresponds to a trait with a perfect narrow-sense
heritability ($h^2 = 1$). Although there are some phenotypes that approach this ideal, it is important to consider the more typical situation of $h^2<1$.
Figure
\ref{fig:rhodelta_random}B shows how the $NE$ varied in the presence of a noise level corresponding
to $h^2=0.5$ (which is roughly the SNP-based heritability of height
  \cite{yang:2010:snpherit,vattikuti:2012}). We can see that the transition boundary was smoothed and effectively
shifted downward.

In the noisy case, the transition boundary was less
dependent on $\delta$ than in the noiseless case. Note that in Figure \ref{fig:rhodelta_random}A-B the noise variance is fixed by $h^2$, but $\rho$ and $\delta$ are both functions of the sample size. Fixing $\rho$
and traversing the phase plane horizontally can be interpreted as
using a sample of size $n$ to study a particular phenotype with $s$
nonzeros, changing the number of genotyped markers in successive assays; Figure~\ref{fig:rhodelta_random}B shows that in the noisy case
an order-of-magnitude change in $p$ had a negligible impact on the
quality of selection.

Given this insensitivity to $\delta$, it is instructive to increase the resolution with which the phase transition can be studied by fixing $\delta$ and then comparing the
$h^2 = 1$ and $h^2 = 0.5$ cases.  Figure \ref{fig:rhodelta_random}C shows that the $NE$
approached its asymptote beyond the theoretical phase transition in
both cases. Moreover, the asymptote appeared to be greater than zero
in the noiseless case. This behavior may suggest that the
noise-dependent $\lambda_{\min}$  prescribed by CS theory is suboptimal when noise is in fact absent, although the closeness of the theoretical and empirical phase
boundaries implies that the deviation from optimality is
mild. The transition was not altered in the noiseless
  case when $\lambda_{\min}$ was estimated using
  cross-validation, although there was some improvement in the noisy case. A 10-fold cross-validation increased the
  computational time by 10 to 100-fold. The similar quality of selection achieved by the theoretical
  $\lambda_{\min}$ and the use of cross-validation supports the theoretical
estimate.

In the noiseless case, when using a criterion of $NE<0.5$, the phase transition to vanishing $NE$ began at $\rho
\approx 0.4$. In the noisy case of $h^2=0.5$, the phase transition began at $\rho \approx
0.03$ ($n \approx 30 s$).  As expected, the sample size for a given number of nonzero
coefficients must be larger in the presence of noise. 

\subsection*{Measures of selection}

We next examined whether nonzeros were being correctly selected despite a nonzero
$NE$ by considering additional measures of selection:
\begin{enumerate}
\item The false positive rate ($FPR$), the fraction of true zero-valued
coefficients that are falsely identified as nonzero.
\item The positive predictive value ($PPV$), the number of correctly selected true nonzeros divided by the total
number of nonzeros returned by the selection
algorithm. $1 - PPV$ equals the false discovery rate ($FDR$).
\item The median of the $P$-values obtained when regressing the phenotype on each of the $L_1$-selected markers in turn ($\mu_\textrm{$P$-value}$). Each such $P$-value is the standard two-tailed probability from the $t$ test of the null hypothesis that a univariate regression coefficient is equal to zero. The
  previous measures of recovery---$NE$, $FPR$, $PPV$---cannot be
  computed in realistic applications because they depend on the unknown
  $\mathbf{x}$, and thus it is of interest to examine whether an
  observable quantity such as $\mu_\textrm{$P$-value}$ also undergoes
  a phase transition at the same critical sample
  size. 
\end{enumerate} 
We hypothesized that a measure of the $P$-value distribution of the putative nonzero set may reflect
 the phase transition since the distribution of $P$-values of normally distributed random variables is
  uniform and is the basis of false discovery
 approaches for the multiple comparisons problem \cite{storey:2003}.

We now turn to the behavior of these performance metrics as a function of sample size. In the
noiseless case (Figure~\ref{fig:nscan_random}A-B), the $NE$ showed a phase transition at $n \approx$ 1,000, but the $PPV$, $FPR$, and $\mu_\textrm{$P$-value}$
converged to zero around $n=$ 1,500. Since we fixed $s$ to
be 125, the location of the transition boundary with respect to the $NE$ at the point
($\rho=0.125$, $\delta=0.125$) was consistent with
Figure \ref{fig:rhodelta_random}A. Also shown is the point ($\rho=0.08$, $\delta=0.19$), where the $PPV$, $FPR$, and $\mu_\textrm{$P$-value}$ converged to zero. 

As the noise was increased (Figure \ref{fig:nscan_random}C), the $NE$ declined less sharply with increasing $n$, as expected from Figure
\ref{fig:rhodelta_random}. In contrast as shown in Figure \ref{fig:nscan_random}D, the other measures (particularly the $PPV$
and $\mu_\textrm{$P$-value}$) neared their asymptotic values even in
the presence of noise. The transitions of
$FPR$, $PPV$, and $\mu_\textrm{$P$-value}$ from poor to good performance were not smoothed by noise to the same extent as the transition of the $NE$.

The greater robustness of the $FPR$, $PPV$ and
$\mu_\textrm{$P$-value}$ against residual variance relative to the
$NE$ shows that accurate \emph{selection} of nonzeros
can occur well before the precise \emph{fitting} of
their coefficient magnitudes. The fact that the observable quantity
$\mu_\textrm{$P$-value}$ exhibits this robustness is particularly
important; a steep decline in $\mu_\textrm{$P$-value}$ across
subsamples of increasing size drawn from a given dataset demonstrates a transition to good recovery and
implies that the full dataset has sufficient power for accurate
identification. This is an empirical finding which deserves further
investigation. 

For $h^2=0.5$ and across all measures of performance other than the $NE$, the transition appeared to be
around $n=$ 5,000. Given $s=125$ and $p=$ 8,027, this corresponds to
($\rho=0.025$, $\delta=0.625$), which is circled in Figure
\ref{fig:rhodelta_random}B. This estimate of the critical $\rho$ is consistent
with our previous estimate when $\delta$ was fixed at 0.5, supporting the
weak dependence on $p$.


\subsection*{Quality of selection in the presence of LD}

We have shown that randomly sampled SNPs from a GWAS of Europeans have the properties of a compressed sensor. This was expected, given that randomly sampled
markers will be mostly uncorrelated and therefore closely approximate
an isotropic matrix. 

We next consider a genotype matrix characterized by LD. To do this while still being able to evaluate recovery at all points of the $\rho-\delta$ plane, we considered all genotyped markers on just chromosome 22. Almost all of these markers were in LD with a few other markers, and the markers within each correlated group
tended to be spatially contiguous (Figure \ref{fig:rhodelta_chr22}C). As
shown in Figures \ref{fig:rhodelta_chr22}A and \ref{fig:rhodelta_chr22}B, the
phase transition boundary with respect to $NE$ was shifted to lower
values of $\rho$ and was less sensitive to $\delta$ as in the noisy case.
  
Although the phase transition from large to small
$NE$ appeared to be affected adversely by LD (at least in the
noiseless case as shown in Figure \ref{fig:rhodelta_chr22}A), the selection
measures were less affected, as seen by comparing Figure
\ref{fig:nscan_chr22} calculated using the intact chromosome 22 with Figure \ref{fig:nscan_random} using markers drawn at random from across the genome. Regardless of LD, the
transition from poor to good values of $\mu_\textrm{$P$-value}$
occurred at nearly the same sample size (about 30 times the number of nonzeros for $h^2=0.5$). 
The $PPV$ and $FPR$ saturated at worse asymptotic
values in the noiseless case. In the noisy case, the
$PPV$ was also lower; perhaps surprisingly, the $FPR$ actually
increased with sample size. 

The relatively poor performance of the $PPV$ and $FPR$ in the case of
LD is somewhat misleading. For example, an ``off-by-one''
(nearby) nonzero called by $L_1$-penalized regression will not count toward the numerator of the $PPV$, even if it is in extremely strong LD with a true nonzero. At the same time, such a near miss does count toward the numerator of the $FPR$. This standard of recovery quality seems overly stringent when we recall that picking out the causal variant from a GWAS ``hit'' region containing multiple marker SNPs in LD continues to be a challenge for the standard MR approach \cite{maller:2012,edwards:2013}.

We examined whether the false positives called by the $L_1$-penalized algorithm were indeed more likely to be in strong LD with the true nonzeros by computing the correlations between
false positives and true nonzeros for $n=$ 5,000 and
$h^2=0.5$. Figure 
\ref{fig:fp_corr1} shows the histogram of the maximum correlation between each false
positive and any of the true nonzeros. We compared this histogram to a realization from the null distribution, generated by drawing markers at random from chromosome 22 and finding each such marker's largest correlation with any of the true nonzeros. The observed histogram featured many more large correlations than the realization from the null
distribution, implying that the false positives showed a significant tendency to be in LD with true nonzeros.

Figure \ref{fig:fp_corr2} provides a visualization of the
correlations among the false positives and true nonzeros. Both false
positives and true nonzeros were sometimes in LD with neighboring
members of their sets; this is to be expected given the short map
length of chromosome 22. The small bursts of elevated color extending
away from the diagonal of the upper-left quadrant suggest that false
positives tend to occur in regions characterized by particularly
strong LD. The striking feature of Figure \ref{fig:fp_corr2} is the
nearly continuous and isolated curve of elevated color slicing through
the off-diagonal blocks of the correlation matrix. This curve shows
that the between-set correlation structure was more complex than a
one-to-one relationship assigning false positives to true
nonzeros. Given the spatial ordering of the SNP indices, the linearity
of the curve demonstrates a marked tendency for called false positives to occur close to one of the true nonzeros.

\subsection*{Sensitivity to the distributions of coefficient magnitudes and MAF}
The appropriate prior on the distribution of coefficient magnitudes is often discussed \cite{gianola:2013:gs}. However, CS
theory shows that the phase boundary for
complete \emph{selection} is relatively insensitive to this
distribution. To test this prediction, we looked for evidence of
performance degradation upon replacing the discrete distribution of
nonzero coefficients used thus far with a hyperexponential
distribution (a mixture of exponential distributions with different
decay constants) (these are defined in Data
Description and shown in Figure~\ref{fig:ensemble}A). The hyperexponential
distribution is a means of implementing an arguably more realistic
ensemble of a few large coefficients followed by a tail of weaker
values \cite{chatterjee:2013}. Figure
\ref{fig:ensemble}B-C shows that, as predicted by theoretical CS
results, for fixed $h^2$ and chromosome 22, the normalized $\mu_\textrm{$P$-value}$ converged to zero at the same sample size regardless of the ensemble.

In the previous simulations, we drew the nonzeros at random from all genotyped markers, thus guaranteeing that the MAF spectra of the nonzeros and the entire genotyping chip would tend to coincide. Here, we also tested whether the MAF spectrum of nonzeros affects the selection phase
boundary. It is known that two SNPs can be in strong LD only if they have similar MAFs
\cite{hedrick:1987,wray:2011}. We confirmed this by taking all pairs
of markers on chromosome 22 and plotting the maximum positive root of the LD measure as a function of squared MAF difference (Figure \ref{fig:maf}A). Therefore, in order to isolate any effect of the MAF distribution among nonzeros not mediated by LD, we constructed a synthetic measurement matrix
$\mathbf{A}$ with independent columns and the same MAF spectrum as
chromosome 22. We then compared recovery
when the nonzero coefficients were sampled from SNPs with
MAF between 0.0045 and 0.015 or when they were sampled above MAF of
0.49. For this we used nonzeros from $\{-1,1\}$. Figure
\ref{fig:maf}B shows no difference in recovery between the conditions
for $h^2=0.5$. This suggests that MAF alone is not a determinant of
the phase transition. Homogeneity in MAF among nonzeros may
enrich correlations as noted above. Such correlations would be expected to reduce the effective
$s$ and thus affect the phase boundary.

\subsection*{Selection of SNPs associated with height}

Motivated by the results above, we examined whether the full sample size of
12,454 subjects was sufficient to achieve the phase transition from poor to good recovery of SNPs associated with a real phenotype (height). We considered the selection measures $\mu_\textrm{$P$-value}$ and adjusted positive predictive value ($PPV^*$); the latter extended true-positive status to any selected SNP within 500 kb of a SNP identified as a likely marker of a height-affecting variant in the
GIANT Consortium's analysis of $\sim$180,000 unrelated individuals
\cite{langoallen:2010}. This extension is consistent with the rule of thumb designating a 1-Mb region as a ``locus'' for purposes of counting the number of GWAS ``hits'' \cite{yang:2012:cond}. The relative insensitivity of $\mu_\textrm{$P$-value}$ to LD suggests that $PPV^*$ rewards the identification of both true nonzeros and markers tagging nonzeros; we therefore substituted $PPV^*$ for $PPV$ in an attempt to align the phase dynamics of our precision measure with those of $\mu_\textrm{$P$-value}$. Whether a selected marker fell within 500 kb of a GIANT-identified marker was determined by consulting the the BROAD SNAP database
(\url{http://www.broadinstitute.org/mpg/snap/}) \cite{johnson:2008:snap}. 

Figure \ref{fig:height1}A shows that $\mu_\textrm{$P$-value}$ failed
to approach zero,  suggesting that that $n=12,454$ is not large enough to see
a phase transition to the regime of good recovery. Given our empirical finding that $\rho \approx 0.03$ is required for $h^2 \approx
0.5$, this suggests that height is affected by at least 400 causal
variants, a result consistent with the observation that the $\sim$250
known height-associated SNPs account for only a small proportion of
this trait's additive genetic variance \cite{yang:2012:cond}. The null
$PPV^*$ derived from randomly chosen SNPs, however, was  smaller than
the observed $PPV^*$ (Figure \ref{fig:height1}A); this was consistent
with the detection of some true signal. In other
  words, although no phase transition was evident, the recovery
  measure did improve with increased sample size.

The penalization parameter $\lambda$ was set using CS theory to
minimize NE error based on the expected noise-level from reported narrow sense heritability for height \cite{yang:2010:snpherit,vattikuti:2012}.  If
$\lambda$ is set too low, then more false positives are expected; if
$\lambda$ is set too high, then true nonzeros will be missed. According to CS theory, an $L_1$-penalized method can still select some of the largest coefficients from a nonuniform distribution of coefficient magnitudes even if complete recovery is out of reach \cite{candes:2006:stable}.
We investigated whether it was possible to achieve a phase transition to low $\mu_\textrm{$P$-value}$ and high $PPV^*$, at the cost of recovering only a small fraction of all true nonzeros, by increasing the penalty parameter
$\lambda$. More specifically, we set $\lambda$ to a higher value consistent with $h^2=0.01$ rather
than 0.5. In this case the $L_1$ algorithm returned 20 putative nonzeros rather
than the original 403, and both
$\mu_\textrm{$P$-value}$ and $PPV^*$ exhibited better performance
(Figure \ref{fig:height1}B). Compared to the less
  stringent $\lambda$, $PPV^*$ as a function of $n$ was less smooth but appeared to stabilize to a high recovery
  value after $\sim 7000$ subjects. Evidently, if the sample size does not suffice to capture the full heritability, setting the penalty parameter to a value appropriate for a lower heritability can lead to a smaller set of selected markers characterized by good precision.
 
Figure \ref{fig:height2} illustrates the physical distances between the markers
selected in our strict-$\lambda$ (assuming $h^2=0.01$) analysis and the markers identified by the GIANT Consortium. Of the 20 $L_1$-selected markers, 14 were within 500-kb of a GIANT-identified marker. However, the $L_1$-selected markers defined to be false positives were still relatively close to
GIANT-identified markers. This may indicate that the 500-kb criterion
for declaring a true positive was too stringent; if so, then our stated
$PPV^*$ of 0.7 can be regarded as a lower bound. As an informal comparison, Figure
\ref{fig:height2} also displays the results of a more standard MR-type
GWAS analysis. For a $P$-value of
$10^{-8}$ and all 12,454 subjects, MR returned six SNPs,  five of which were GIANT-identified markers, and four were exact matches with SNPs selected by our $L_1$ algorithm (Figure \ref{fig:height2}). With a $P$-value cutoff of $5\times 10^{-8}$ and all subjects, MR returned 13 markers, 10 of which were GIANT-identified, and 7 of which were identical to the $L_1$-selected markers.

The presence of a phase transition is not necessarily restricted to $L_1$
algorithms but rather may represent a deeper phenomenon in signal
recovery. Other methods may show a similar phase transition---although
CS theory suggests that, among convex optimization methods, those within the $L_1$ class are
closest to the optimal combinatorial $L_0$ search. 
We conducted additional analyses to test whether a phase transition at
a critical sample size could also be observed when our height data
were analyzed using the MR approach commonly used in GWAS. In these
simulations we varied the $P$-value threshold for genome-wide
significance. As measures of selection are potentially subject to a phase transition, we examined the $PPV^*$ and the adjusted median
$P$-value ($\mu_\textrm{$P$-value}^*$). The latter measure was defined to be the median $P$-value among those SNPs surviving the $P$-value cutoff, divided by the cutoff itself; the normalization was necessary to remove the dependence on the choice of cutoff. As shown in Figure
\ref{fig:assoc}, the $P$-value threshold $10^{-8}$ yielded very few selected SNPs, and in fact none were returned at sample sizes smaller than about 8,000.
 However, $\mu_\textrm{$P$-value}^*$
was mostly close to zero in the region of Figure \ref{fig:assoc}B corresponding to $n > 8,000$ and $ P\textrm{-value} < 10^{-6}$, suggesting that true
nonzeros were being selected.  This is confirmed by the fact that the $PPV^*$ typically exceeded 0.6 in this same region (Figure \ref{fig:assoc}A). For
$P$-value thresholds less stringent than $10^{-6}$, signs of a phase transition at a critical sample size were still discernible. 

A search for a phase transition can be a useful approach to
determining the optimal $P$-value threshold in standard GWAS protocols
employing MR. In addition to \emph{a priori} assumptions regarding the
likely number of true nonzeros and their coefficient magnitudes
\cite{wtccc:2007,chatterjee:2013} and agreement between studies of
different designs \cite{turchin:2012}, GWAS investigators might rely
on whether a measure such as $\mu_\textrm{$P$-value}^*$ undergoes a
clear phase transition as they take increasingly large subsamples of
their data. A majority of markers surviving the most liberal
significance threshold bounding the second phase are likely to be true
positives. 

\section*{Discussion}

Our results with real European GWAS data and simulated vectors of regression coefficients demonstrate the accurate selection of those markers with nonzero coefficients, consistent with CS sample size requirements ($n$) for a given sparsity ($s$) and total number of predictors ($p$). We found that the matrix of standardized genotypes
exhibits the theoretical phase transition between poor
and complete selection of nonzeros (Proposition~\ref{donoho_theorem}). We also found, as for Gaussian random matrices in earlier
studies, that the phase transition depends on the scaling ratios
$\rho = s/n$ and $\delta = n/p$ \cite{donoho:2010:precise}.

We obtained results regarding
the effect of noise (i.e., $h^2 < 1$) that are consistent with
earlier empirical studies of random
matrices and recently proven theorems \cite{donoho:2006:break,donoho:2011,candes:2011:ripless}. Roughly speaking, we show that the
critical sample size is determined mainly by the ratio
  of $s$ to $n$ and only weakly sensitive to $p$,
particularly as noise increases. For example, if $h^2=0.5$, which is
roughly the narrow-sense heritability of height and a number of other
quantitative traits
\cite{yang:2010:snpherit,davies:2011,vattikuti:2012}, we find that $\rho$ should be less than approximately
0.03 for recovery
irrespective of $\delta$. There is no hope of recovering the complete vector of coefficients $\mathbf{x}$ above this threshold (i.e. smaller sample sizes). For example, if we have prior knowledge that
$s=1,200$, then this means that the sample size should be no less than 40,000
subjects.  We find empirically that for $h^2 \sim 0.5$, $n \sim
30 s$ is sufficient for selection of the nonzeros.

In real problems we cannot rely on measures of model recovery based on
the unknown $\mathbf{x}$. Hence, we introduced a new measure based on the median $P$-value
of the $L_1$-selected nonzeros, $\mu_\textrm{$P$-value}$. We found that $\mu_\textrm{$P$-value}$ provides a robust means of detecting the boundary between poor and good recovery. Proposition~\ref{candes_theorem} shows that the
recovery error $NE$ in the favorable phase scales with $\rho$ and noise; however, we
observed that the recovery measures $FPR$, $PPV$ and $\mu_\textrm{$P$-value}$ approached zero
faster than the $NE$, confirming that accurate identification of
nonzeros can occur well before precise estimation of their
magnitudes. 


An $L_1$-penalized regression algorithm is equivalent to linear regression with a Laplace prior distribution of coefficients, and in theory a Bayesian method invoking a prior distribution better matching the unknown true distribution of nonzero coefficients should outperform the lasso in effect estimation. However,  it is by no means clear that the performance of $L_1$ penalization with respect to selection can be bettered. 
For example, the lasso and BayesB display rather similar performance properties \cite{wimmer:2013}. However, both methods clearly outperformed ridge regression (a
non-$L_1$ method), which exhibited no phase transition away from poor
performance. 
Furthermore, it is usually accepted by GWAS researchers that knowledge of the markers with nonzero coefficients may be quite valuable, even if the
actual magnitudes of the coefficients are not well determined. Combining the advantages of different approaches by applying one of them to the $L_1$-selected markers is a possibility.

Perhaps contrary to intuition, but consistent with theoretical results for CS \cite{donoho:2009,donoho:2010:precise}, we found
 that the phase transition to good recovery (at least as measured by $\mu_\textrm{$P$-value}$) was insensitive to 
the distribution of coefficient magnitudes. It is well known in CS that $L_1$-penalized regression is
  nearly minimax optimal (minimizes the error of the worst case) and that the phase transition is robust to the
  distribution of coefficient magnitudes. In some cases a good prior may reduce the
  mean-square error and shift the location of the phase
  transition \cite{vila:2013}. Simulations supporting this notion, however, were performed with much
  higher signal-to-noise ratio (SNR) than hypothesized for realistic
  GWAS problems. The performance increase was attenuated as the
  SNR was decreased to levels still higher than usual in  GWAS (10dB or
  $h^2>0.9$ where SNR on the dB scale is given by $10 \cdot \log_{10} \left(
    \frac{\sigma^2_A}{\sigma^2_E} \right)$. These algorithms are currently being
  explored in lower-SNR regimes.
 We observed that cross-validation did slightly affect the phase transition boundary in the
  noisy case; nevertheless the theoretical penalization parameter proved to be a good
  ``rule of thumb'' for initial screening. Calculating the theoretical penalty depends on
  knowledge of $h^2$, which may be estimated using the genomic-relatedness method
  \cite{yang:2010:snpherit,vattikuti:2012,vattikuti:2014,lee:2014}.

%

Genomic selection methods have been criticized by researchers who doubt that the number of nonzeros ($s$) will typically be smaller than a practically attainable sample size ($n$) \cite{gianola:2013:gs}. The application of CS theory circumvents this problem because it allows the optimization method to self-determine whether or not the nonzero markers are sufficiently sparse compared to the sample size.  No prior assumptions are required.
Furthermore, in humans there is evidence that a
number of traits satisfy the sparsity assumption, at least with
respect to common variants contributing to heritability
\cite{park:2011,stahl:2012,ripke:2013}.  

CS theory does not provide performance guarantees in the presence of
arbitrary correlations (LD) among predictor variables: it must be
verified empirically, as we have done.  In agreement with previous results \cite{wimmer:2013}, 
we find that the phase transition as measured by NE is strongly affected by LD.  However,
according to our simulations
using all genotyped SNPs on chromosome 22, $L_1$-penalized regression
does select SNPs in close proximity to true nonzeros. The difficulty
of fine-mapping an association signal to the actual causal variant is
a limitation shared by all statistical gene-mapping
approaches---including marginal regression as implemented in standard
GWAS---and thus should not be interpreted as a drawback of $L_1$
methods.

We found that a sample size of 12,464 was not sufficient to achieve
full recovery of the nonzeros with respect to height. The penalization
parameter $\lambda$, however, is set by CS theory so as to minimize
the $NE$ based on the expected noise-level. In some situations it might be desirable to tolerate a relatively large $NE$ in order to achieve precise but incomplete recovery (few false positives, many false negatives). By setting $\lambda$ to a strict value appropriate for a low-heritability trait (in effect, looking for a subset of markers that account for only a fraction of the total heritability, with consequently higher noise), we found that a phase transition to good recovery can be achieved with smaller sample
sizes, at the cost of selecting a smaller number of markers and hence suffering many false negatives.

One interesting feature of the recovery measure based on the median
$P$-value ($\mu_\textrm{$P$-value}$) is that it seemed to rise as the
sample size was increased in the region of poor recovery and then fall
after the sample size crossed the CS-determined phase transition
boundary. This rise and then fall was very dramatic in our simulations
(Figures \ref{fig:nscan_random} and \ref{fig:nscan_chr22}) and also appeared in our analysis of height
(Figure \ref{fig:height1}). This behavior may be a consequence of the
fact that as the sample size is increased, $\lambda$ in the algorithm is decreased
(see Methods). Hence, in the region of poor recovery, the relaxation of
the penalty with increasing sample size may permit the selection of
more SNPs and hence the inflation of the $FPR$ and
$\mu_\textrm{$P$-value}$. However, once the phase transition to good
performance begins, the recovery measures begin their characteristic
sharp decrease. This non-monotone behavior accentuates the transition
boundary and can be exploited to aid its detection. 

In summary, compressed sensing utilizes properties of
high-dimensional systems that are surprising from the perspective of classical statistics. The regression problem faced by GWAS and GS is
well-suited to such an approach, and we have shown that the matrix of
SNP genotypes formed from European GWAS data is in fact a
well-conditioned sensing matrix. Consequently, we have inferred the sample sizes required to achieve accurate
model recovery and demonstrated a method for determining whether the minimal sample size has in fact been obtained.

\section*{Methods}

\subsection*{$L_1$-penalized regression algorithm}

$L_1$-penalized regression (e.g., lasso) minimizes the objective function
\begin{equation}\label{lambda1}
\|\hat{\mathbf{y}}- \mathbf{y}\|_{L_2}^2+\lambda\|\hat{\mathbf{x}}\|_{L_1}
\end{equation}
where $\hat{\mathbf{y}}$ is the estimated breeding value given by
$\mathbf{A}\hat{\mathbf{x}}$. The setting of the penalization parameter $\lambda$ is described below.

The algorithm was performed using pathwise coordinate optimization and the
soft-threshold rule \cite{friedman:2007}. Regression coefficients were sequentially updated with
\begin{equation}\label{lasso_path}
 \hat {\mathbf{x}}_j(\lambda)\leftarrow S \left(\hat{\mathbf{x}}_j(\lambda)+\frac{1}{n}\sum\limits_{i=1}^{n} \mathbf{A}_{ij}(\mathbf{y}_i-\hat{\mathbf{y}}_i),\lambda\right)\textrm{ for } j=1,2,...,p
\end{equation}
where
\begin{eqnarray}
 S \left(z,\lambda\right) &\equiv& \textrm{sign}(z)(|z|-\lambda)_+ \nonumber\label{softthreshold1}\nonumber\\
\label{softthreshold2}
&=&\begin{cases}
z-\lambda, & \mbox{if } z>0 \mbox{ and } \lambda<|z|,\\
z+\lambda, & \mbox{if } z<0 \mbox{ and } \lambda<|z|,\\
0, & \mbox{if } \lambda\geq|z|\\
\end{cases}
\end{eqnarray}

We assumed convergence if the fractional change in the objective function given by Equation~\ref{lambda1} was less than $10^{-4}$. In addition, we performed
lasso with a warm start \cite{friedman:2010},
using a logarithmic descent of 100 steps in $\lambda$ with
$\lambda_\textrm{max} =  (1/n)\|\mathbf{A}'\mathbf{y}\|_{L_\infty}$. For $\lambda_\textrm{min}$
we used $(\sigma_E^*/n)\|\mathbf{A}'\mathbf{e}\|_{L_\infty}$, where $\sigma_E^*=\sqrt{\sigma_E^2+1/n}$
\cite{candes:2011:ripless}. To estimate $\| \mathbf{A}^\prime
\mathbf{e} \|_{L_\infty}$ we created 1,000 sample vectors of $\mathbf{e}$, each
constructed with $n$ i.i.d. normal elements with
 mean zero and variance one, and took the median across
samples of $\| \mathbf{A}^\prime \mathbf{e} \|_{L_\infty}$ scaled by
$\sigma_E^*$.  Estimates of $(\sigma^2_A, \sigma^2_E)$ with
respect to the variants assayed in a given study can be obtained
using the genomic-relatedness method
\cite{yang:2010:snpherit,vattikuti:2012,vattikuti:2014,lee:2014}.  The algorithm can also accommodate any other covariates.

\subsection*{Platform}
Simulations and analyses were performed using MATLAB 2013 (The MathWorks Inc., Natick, Massachusetts) and PLINK
2 (\url{https://www.cog-genomics.org/plink2/}) \cite{purcell:2007}. The $L_1$-optimization algorithm was written in 
MATLAB (available at \url{https://github.com/ShashaankV}) and also a feature of PLINK 2. $P$-values were estimated using MATLAB's  \emph
{regstats} function and PLINK 2. Color-coded phase plane figures
were generated by sampling the $\rho-\delta$ plane and
interpolating between points using MATLAB's
\emph{scatteredInterpolant} function.  GWAS
data were obtained from dbGaP as described in Data Description.  Simulated mock data and the
analysis scripts are available and maintained at \url{https://github.com/ShashaankV}.

\subsection*{Statistics}

The normalized coefficient error ($NE$) is
\begin{equation}\label{ne}
\dfrac{\|x-\hat{x}\|_{L_2}}{\|x\|_{L_2}}.
\end{equation}

The false positive rate ($FPR$) is the fraction of true zero-valued
coefficients that are falsely identified as nonzero.
The positive predictive value ($PPV$) is  the number of correctly selected true nonzeros divided by the total
number of nonzeros returned by the selection
algorithm. $1 - PPV$ equals the false discovery rate ($FDR$).
%
The adjusted positive predictive value ($PPV^*$)
is similar to the standard $PPV$, except that any selected nonzero coefficient
falling within 500 kb of a GIANT-identified marker is counted as a true
positive \cite{langoallen:2010}. 

The median of the $P$-values for the set of putative
  nonzeros ($\mu_{P-value}$) is obtained by: 1) regressing the phenotype
  on each of the $L_1$-selected markers in turn, 2) estimating each $P$-value
  as the standard two-tailed probability from the $t$ test of the null
  hypothesis that a univariate regression coefficient is equal to
  zero, and 3) taking the median over the independent tests. This procedure is independent of the selection algorithm and
  calculated after the $L_1$-penalized algorithm has converged. The adjusted median $P$-value
($\mu_\textrm{$P$-value}^*$) is the median of the MR $P$-values falling below the significance
threshold divided by the threshold itself.

The LD measure ($r^2$) is the squared estimate of the
Pearson's product-moment correlation between the standardized
zero-mean, unit-variance SNPs.

As noted above the raw data is available through dbGaP. Mock data and the
analysis codes are available and maintained at \url{https://github.com/ShashaankV}.


\begin{backmatter}

\section*{Abbreviations}
ARIC: Atherosclerosis Risk in Community; CS: compressed sensing; FDR: false discovery rate; FPR: false positive rate; GENEVA: Gene Environment Association Studies; GIANT: Genetic Investigation of Anthropometric Traits; GS: genomic selection; GWAS: genome-wide association study; LD: linkage disequilibrium; LE: linkage equilibrium; MAF: minor allele frequency; MR: marginal regression; NE: normalized error; OLS: ordinary least squares; PPV: positive predictive value; SNP: single-nucleotide polymorphism.

\section*{Competing interests}
  The authors declare that they have no competing interests.

\section*{Author's contributions}
SV performed the numerical experiments and analyzed the data. SV, JJL,
SDHH, and Chow contributed to the conception of the study, drafted the
article, and endorsed the final version for submission. Chang ported the
 MATLAB $L_1$-penalized regression codes to PLINK 2 for use
in the height analysis.

\section*{Acknowledgments}
We thank Nick Patterson for comments on earlier versions
  of this work and Phil Schniter for input on the EM-GM-AMP
  algorithm. This work was supported by the Intramural Program of the
  NIH, National Institute of Diabetes and Digestive and Kidney
  Diseases (NIDDK). 

The Atherosclerosis Risk in Communities Study is carried out as a collaborative study 
supported by National Heart, Lung, and Blood Institute contracts (HHSN268201100005C, 
HHSN268201100006C, HHSN268201100007C, HHSN268201100008C, 
HHSN268201100009C, HHSN268201100010C, HHSN268201100011C, and 
HHSN268201100012C). The authors thank the staff and participants of the ARIC study for their 
important contributions. Funding for GENEVA was 
provided by National Human Genome Research Institute grant U01HG004402 (E. Boerwinkle).
  

\bibliographystyle{bmc-mathphys} 
\bibliography{ms_gigascience_references_revised}      



\newpage
\section*{Figures}

\begin{figure}[h!]
 \centering
   \includegraphics[width=1.0\textwidth]{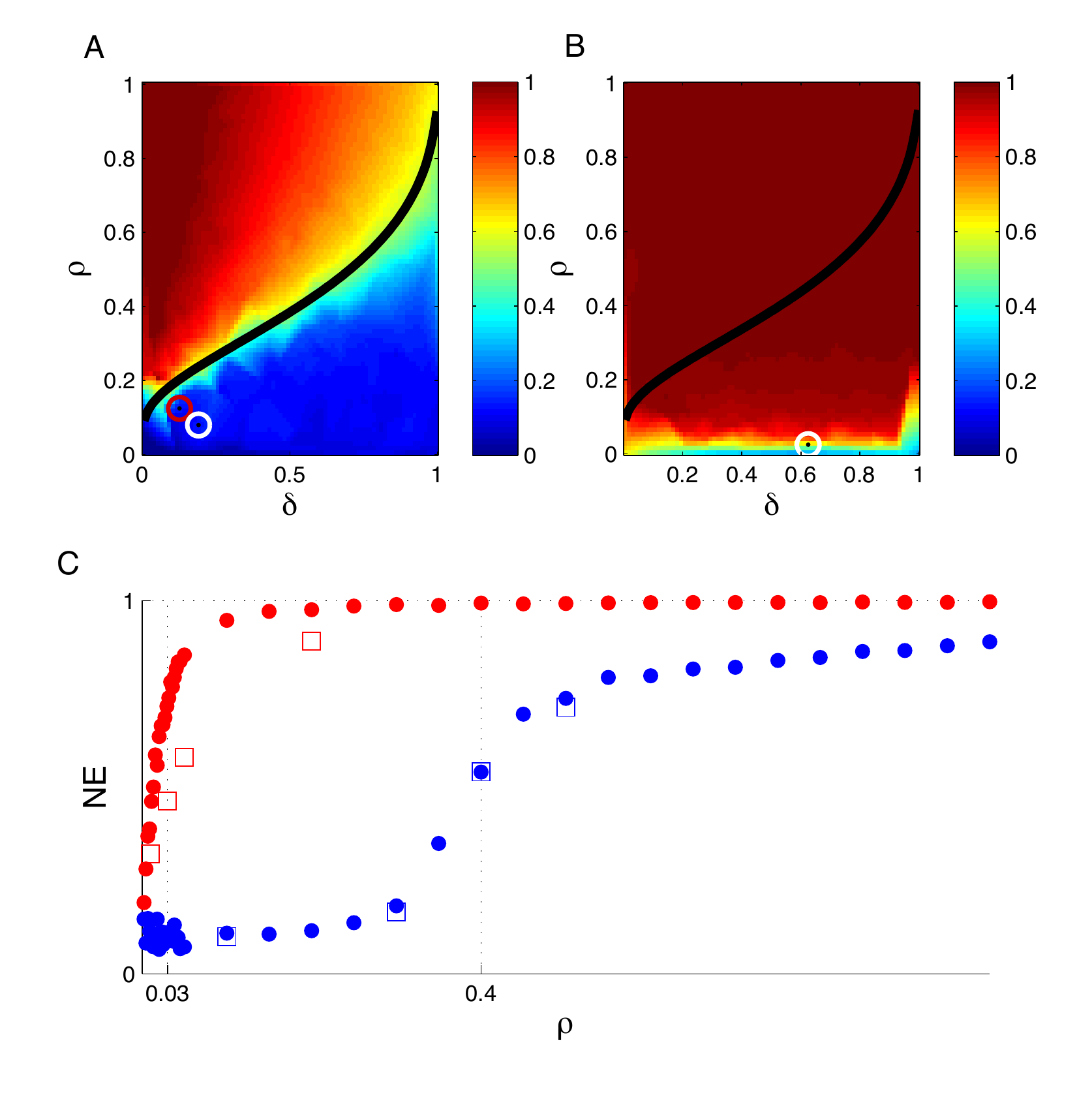}
    \caption{\textbf{Error in the $\rho-\delta$ plane for a
        measurement matrix of random genomic SNPs ($\rho=s/n$ and $\delta=n/p$).} \textbf{(A)} Color corresponds to the normalized
      error ($NE$) of the coefficients
      $\dfrac{\|x-\hat{x}\|_{L_2}}{\|x\|_{L_2}}$. The black curve is the
      expected phase boundary between poor and good recovery from
      \cite{donoho:2006:neighbor}. The number of SNPs, $p$, was fixed
      at 8,027. The heritability was set to one (noiseless case). The circles correspond to the points ($\rho=0.08$, $\delta=0.19$) (white) and ($\rho=0.125$, $\delta=0.125$)
   (red) discussed in \emph{Measures of selection}. \textbf{(B)} Same as panel (A), except that the heritability was set to 0.5 (noisy case). The white circle corresponds to the point ($\rho=0.025$, $\delta=0.625$) discussed in
  \emph{Measures of selection}. \textbf{(C)} $NE$ versus $\rho$
      for fixed $n=$ 4,000 and $p=$ 8,027 (blue corresponds to $h^2 =
      1$, red to $h^2 = 0.5$). The square markers indicate recovery quality
      evaluated at a few data points using the lasso algorithm with 10-fold cross-validation written
      by MATLAB.}
      \label{fig:rhodelta_random}
\end{figure}

\begin{figure}[h!]
 \centering
 \includegraphics[width=1.0\textwidth]{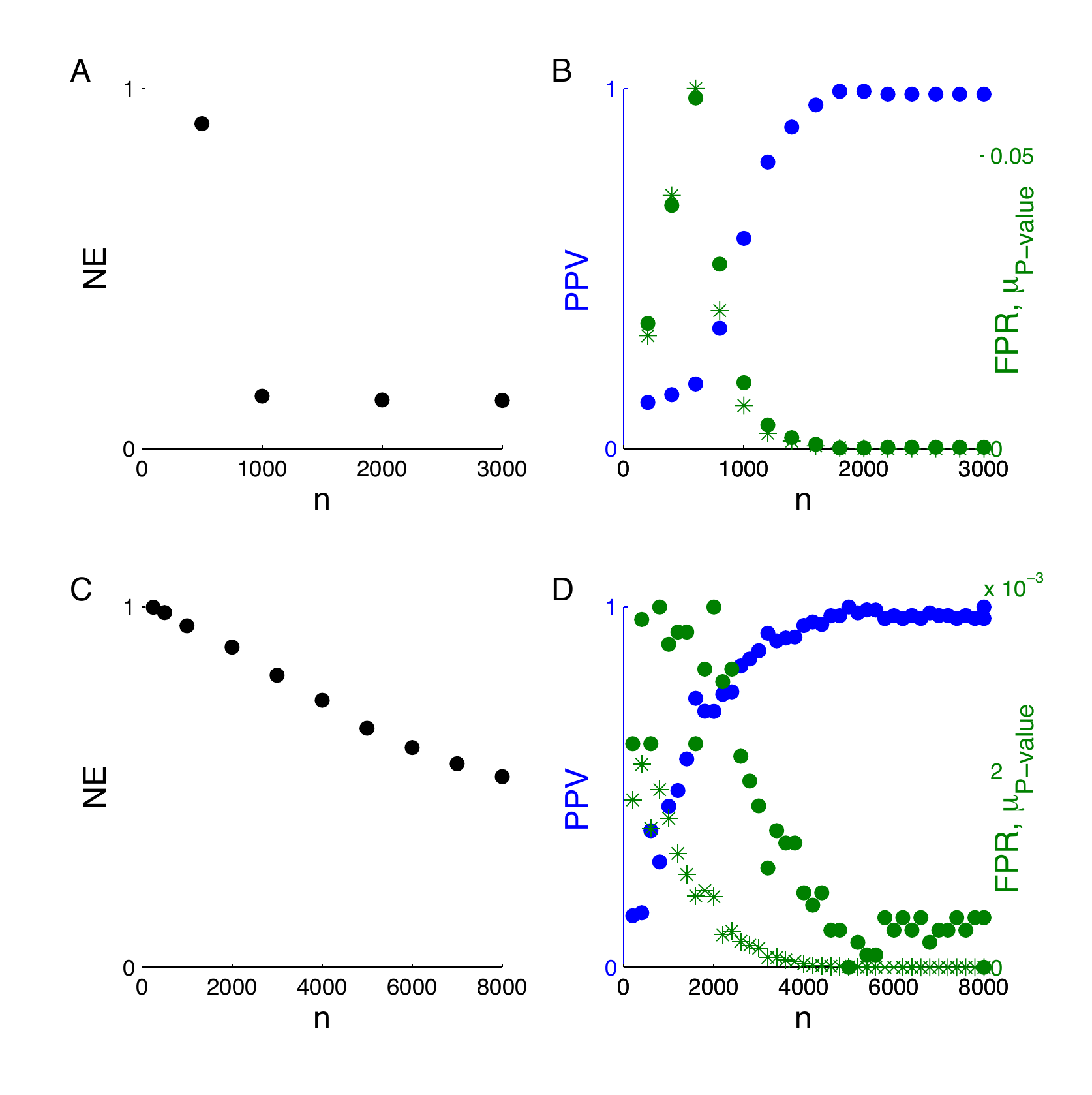}
    \caption{\textbf{Measures of selection as a function of
        sample size for the measurement matrix of random genomic SNPs.} Fixing $s=125$
      and $p=$ 8,027, we measured the selection of true
      nonzero coefficients according to four metrics for $h^2=1$
      \textbf{(A-B)} and $h^2=0.5$ \textbf{(C-D)}. Shown in
      \textbf{(A-C)} is the normalized error of the coefficients
      ($NE$). Shown in \textbf{(B-D)} are the positive predictive value
      ($PPV$, blue dots), false positive rate ($FPR$, green dots), and median $P$-value
      ($\mu_\textrm{$P$-value}$, green asterisks). The point $n=1,000$
    corresponds to ($\rho=0.125$, $\delta=0.125$) and $n=5,000$ to
    ($\rho=0.025$, $\delta=0.625$) noted in Figures
    \ref{fig:rhodelta_random} A and B respectively.} \label{fig:nscan_random}
\end{figure}

\begin{figure}[h!]
 \centering
  \includegraphics[width=1.0\textwidth]{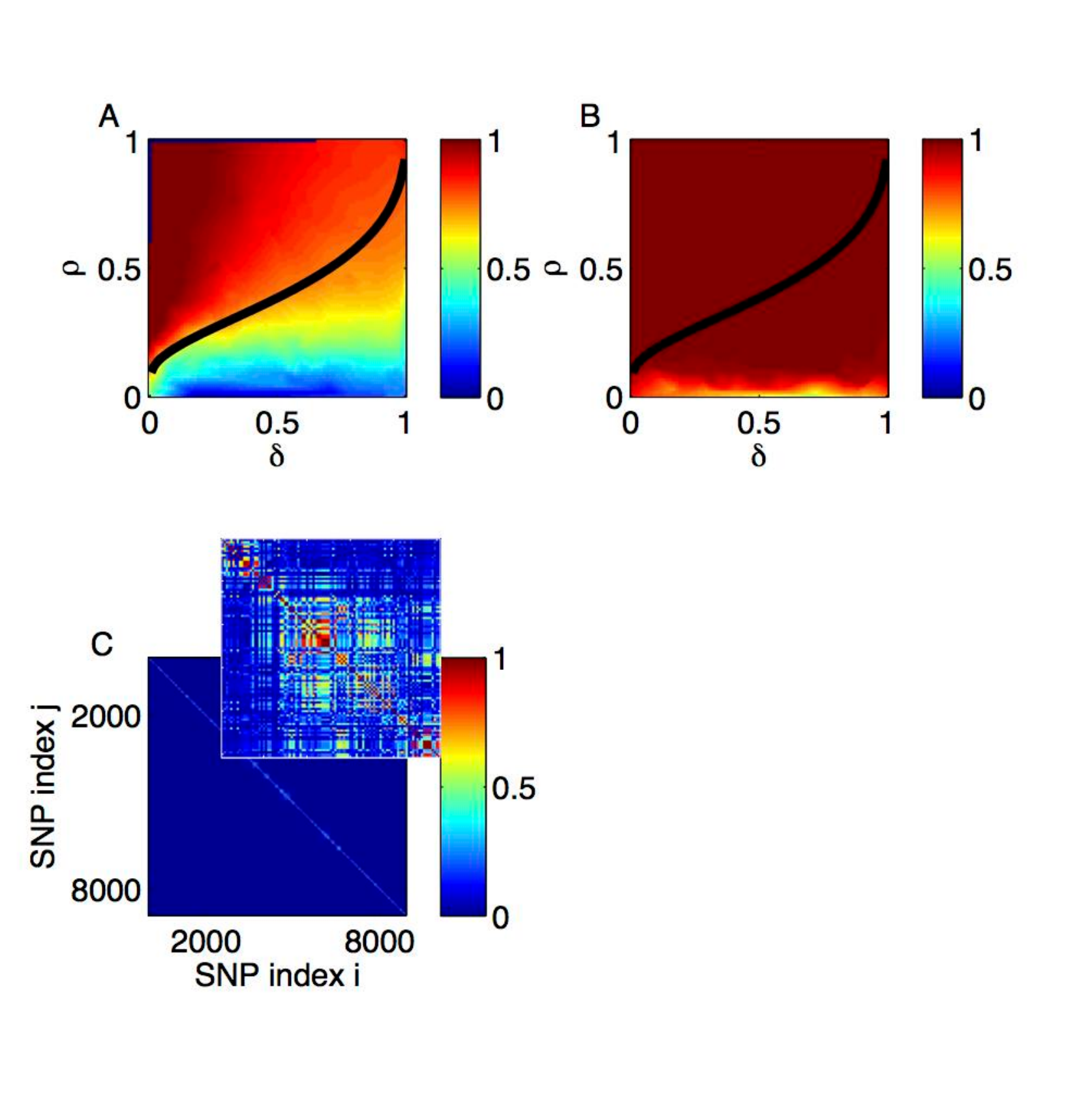}
    \caption{\textbf{Analysis of chromosome 22.} \textbf{(A)} The $\rho-\delta$ plane
      for $h^2 = 1$. $p$ was set to 8,915. Superimposed is the expected phase boundary when there is neither noise nor LD. \textbf{(B)} The same as panel (A), except for $h^2 = 0.5$. \textbf{(C)} The matrix of correlations (positive roots of the $r^2$ LD measure) between genotyped SNPs on chromosome 22. Inset is a $100 \times 100$ sample along the
     diagonal.}
     \label{fig:rhodelta_chr22}
\end{figure}

\begin{figure}[h!]
 \centering
  \includegraphics[width=1.0\textwidth]{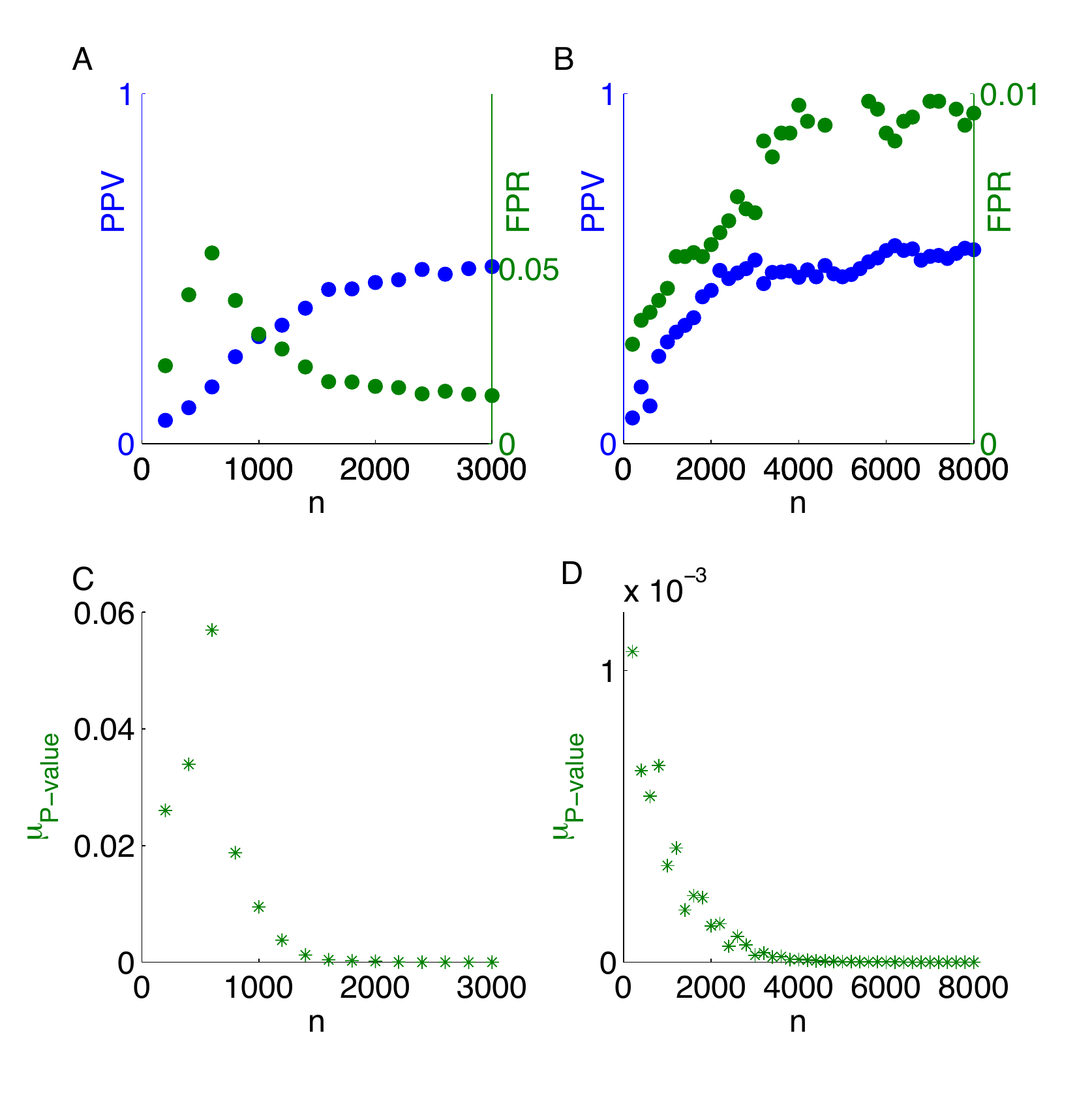}
    \caption{\textbf{Measures of selection as a function of
        sample size for chromosome 22 ($s=125$ and $p=8,915$).} The PPV
      (blue) and FPR (green) for $h^2=1$ \textbf{(A)} and $h^2=0.5$
      \textbf{(B)}. $\mu_{P-value}$ for $h^2=1$ \textbf{(C)} and $h^2=0.5$ \textbf{(D)}. }\label{fig:nscan_chr22}
\end{figure}

\begin{figure}[h!]
 \centering
\includegraphics[width=1.0\textwidth]{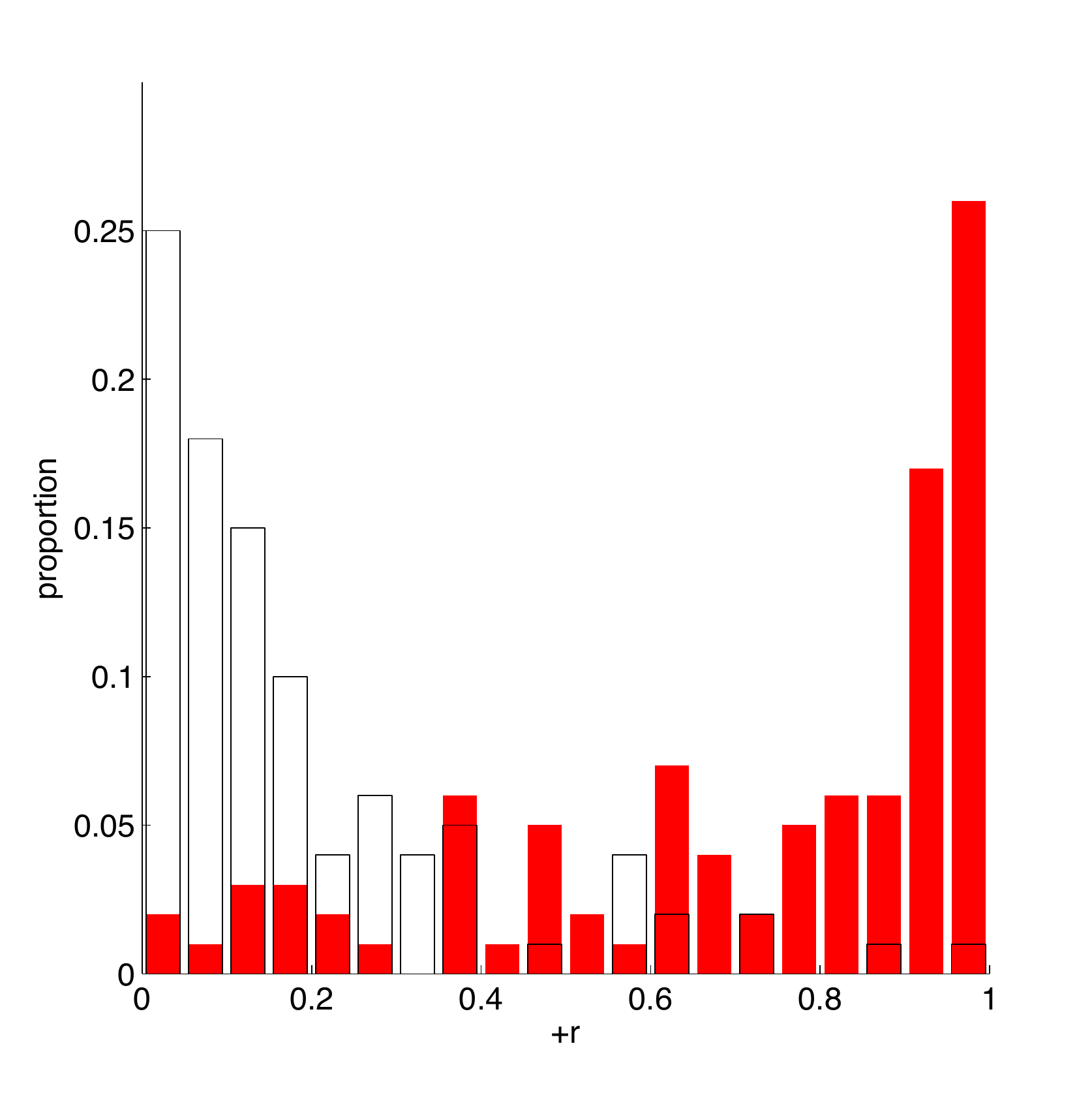}
    \caption{\textbf{Distribution of maximum correlations between
        false positives and
        true nonzeros after the presumptive $\mu_{P-value}$ phase
        transition for chromosome 22.} Histogram of the
    maximum correlation (maximum of the positive roots of the $r^2$ LD measure) between a false positive and true
    nonzero for chromosome 22, given $s=125$, $n=$ 5,000, and
    $h^2=0.5$ (red). Also shown is one realization from the null
    distribution, generated by drawing an equal number of ``false positives'' at random from chromosome 22 (white).}\label{fig:fp_corr1}
\end{figure}

\begin{figure}[h!]
 \centering
\includegraphics[width=1.0\textwidth]{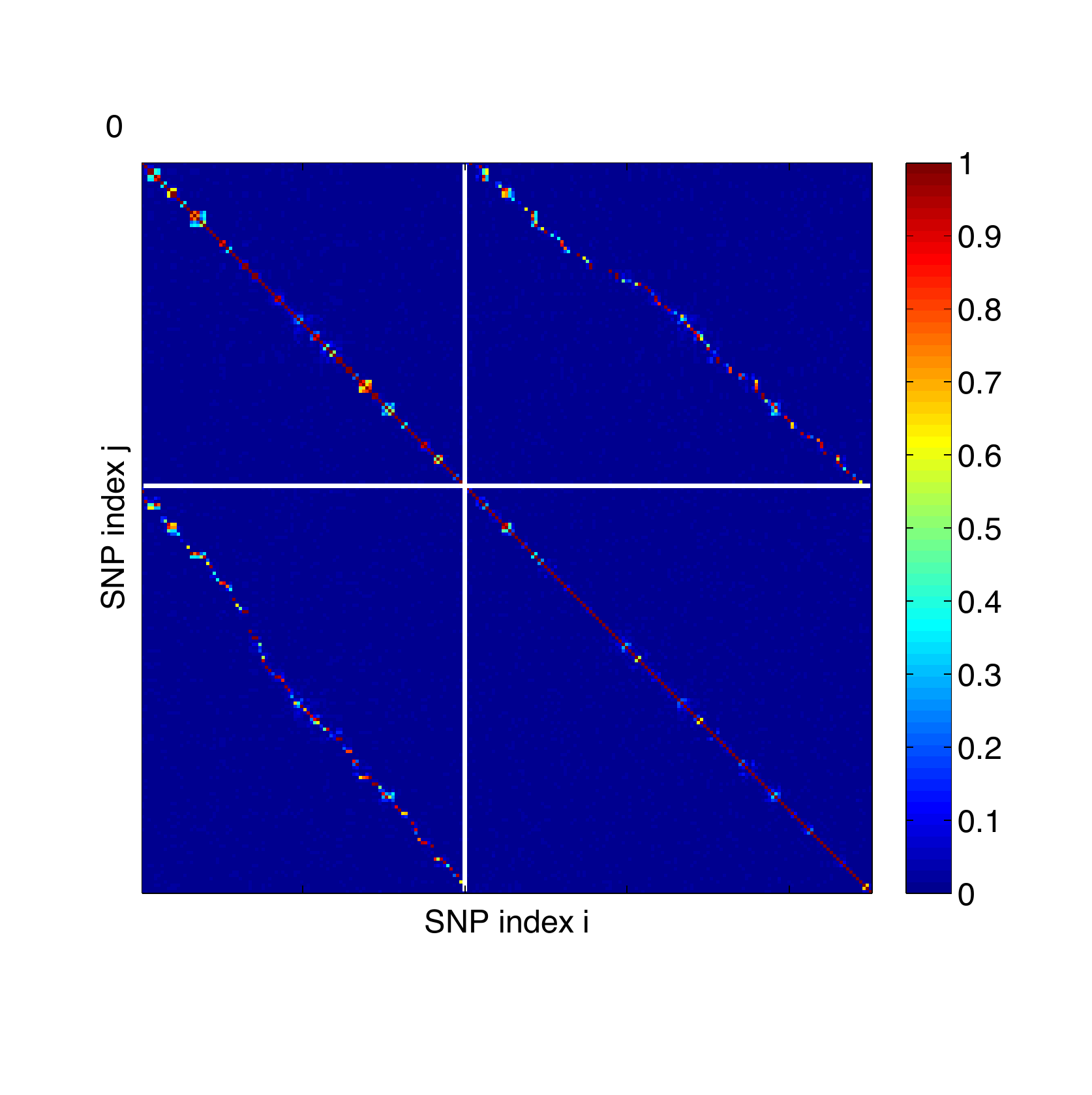}
    \caption{\textbf{The matrix of correlations (positive roots of the
        $r^2$ LD measure) among false positives and true nonzeros after the presumptive $\mu_{P-value}$ phase
        transition for chromosome 22 ($s=125$, $n=5,000$, and $h^2=0.5$).} SNP indices
      begin at the top left corner. The upper-left quadrant contains the correlations among false positives and the lower-right quadrant contains the correlations among the true nonzeros. Each element in the upper-right (lower-left) quadrant represents a correlation between a false positive
    and a true nonzero. Within both the false positive and the true
    nonzero sets, the markers are arranged in order of chromosomal map position.}\label{fig:fp_corr2}
\end{figure}

\begin{figure}[h!]
 \centering
\includegraphics[width=1.0\textwidth]{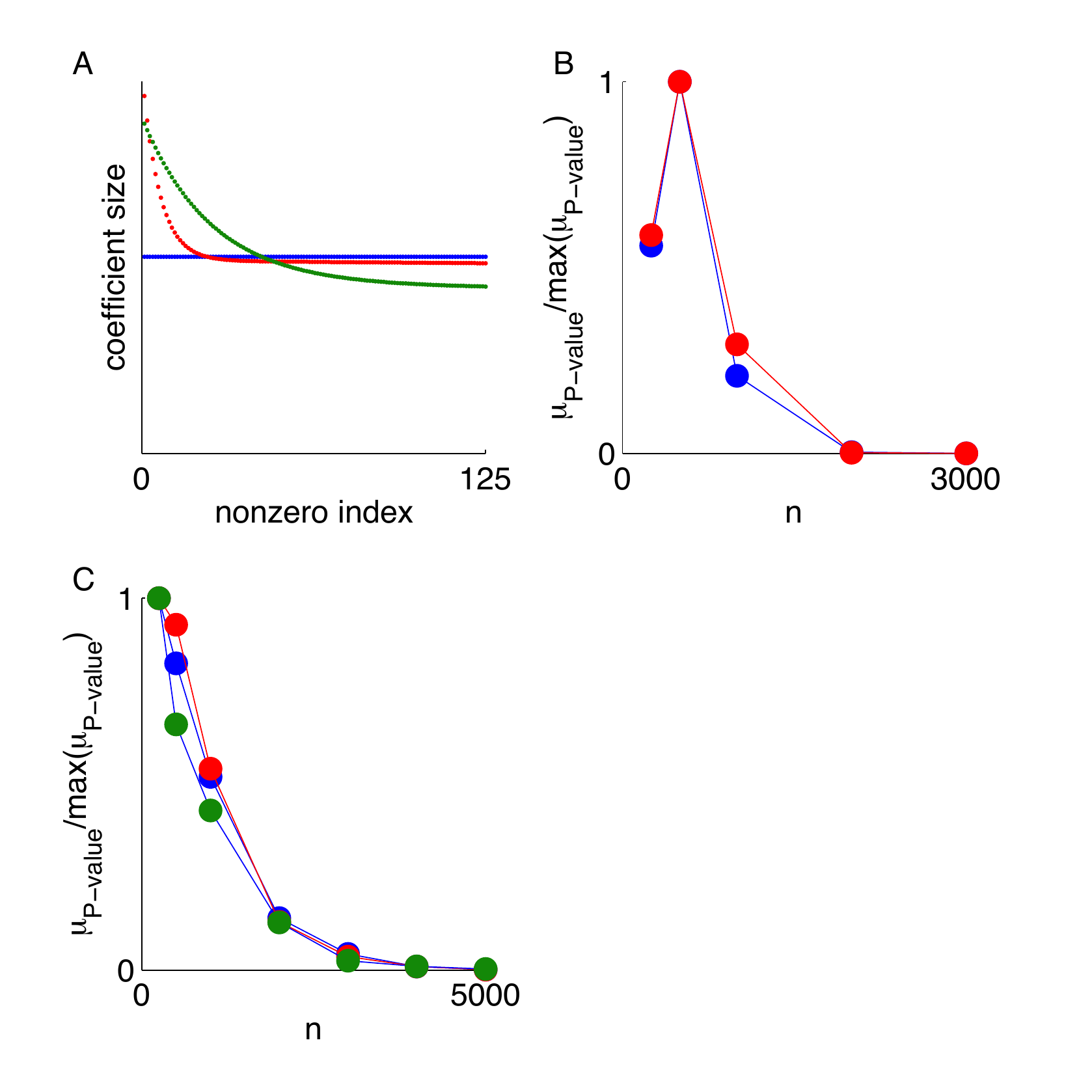}
    \caption{\textbf{Insensitivity of the selection phase boundary to the distribution of coefficient magnitudes
      (ensemble).} \textbf{(A)} $s = 125$ coefficient magnitudes (``effect sizes'') ordered from large to small for the Uniform
    (blue), Hyperexponential 1 (red), and  Hyperexponential 2 (green)
    ensembles. \textbf{(B)} Chromosome 22 analysis using
    $\mu_\textrm{$P$-value}$ to measure selection (normalized by the maximum value) as a
    function of sample size for $h^2=1$ for the Uniform
    (blue) and Hyperexponential 1 (red) ensembles. \textbf{(C)} As in
panel (B) except for $h^2=0.5$. Also shown is recovery for the
    Hyperexponential 2 ensemble (green).}\label{fig:ensemble}
\end{figure}

{\color{red}
\begin{figure}[h!]
 \centering
\includegraphics[width=1.0\textwidth]{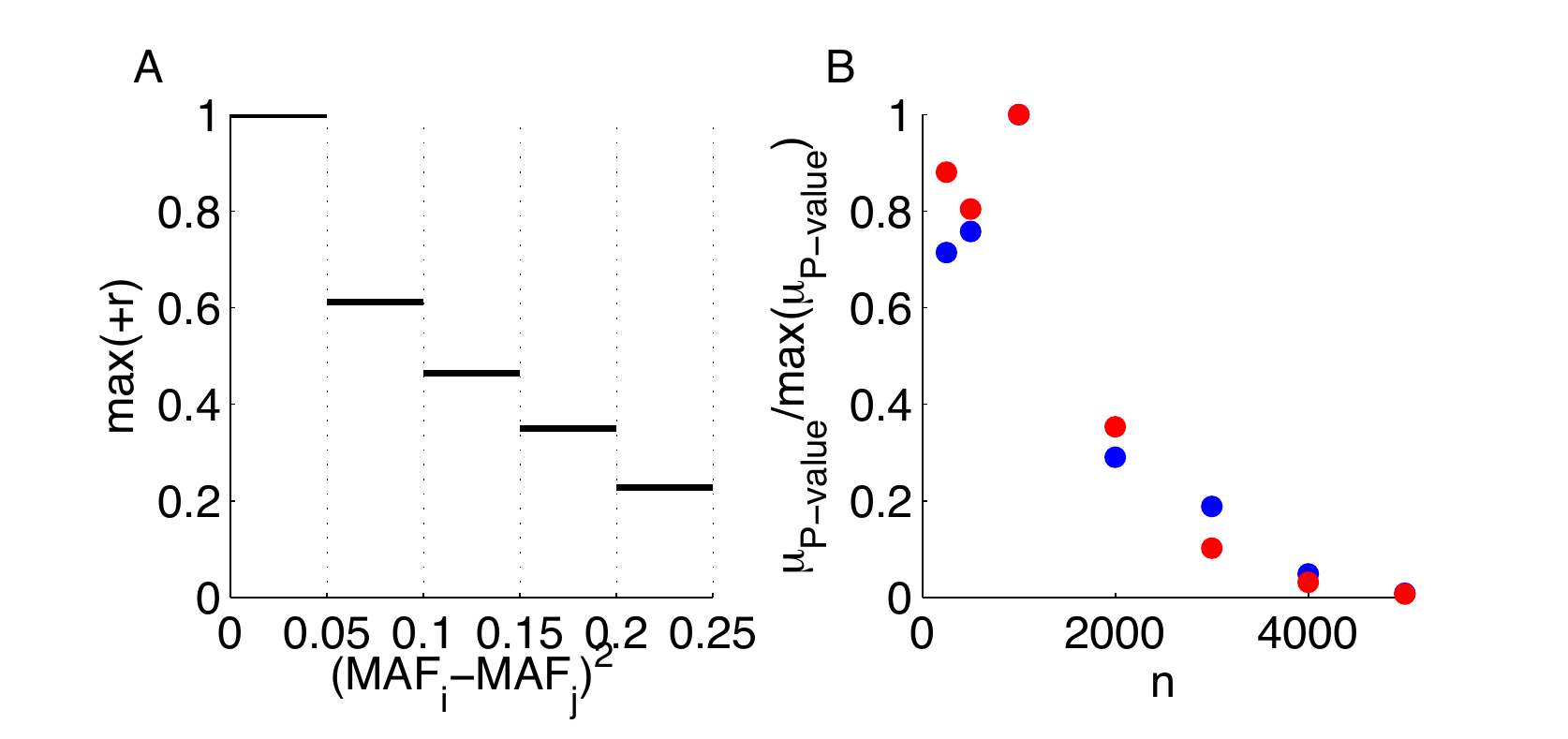}
    \caption{\textbf{Insensitivity of the selection phase
          boundary to minor allele frequency (MAF) for chromosome 22.} \textbf{(A)} The
        maximum positive root of the $r^2$ LD measure (+r) as a function of squared
        MAF difference. The maxima are estimated over bin lengths of
        0.05 for SNPs in chromosome 22. \textbf{(B)} The median
    $P$-value ($\mu_\textrm{$P$-value}$) normalized by the maximum value as a
    function of sample size for $s=125$ from $\{-1,1\}$ and $h^2=0.5$ for nonzero coefficients
    sampled from low (blue) or high (red) MAF SNPs on chromosome 22.}\label{fig:maf}
\end{figure}
}

\begin{figure}[h!]
 \centering
 \includegraphics[width=1.0\textwidth]{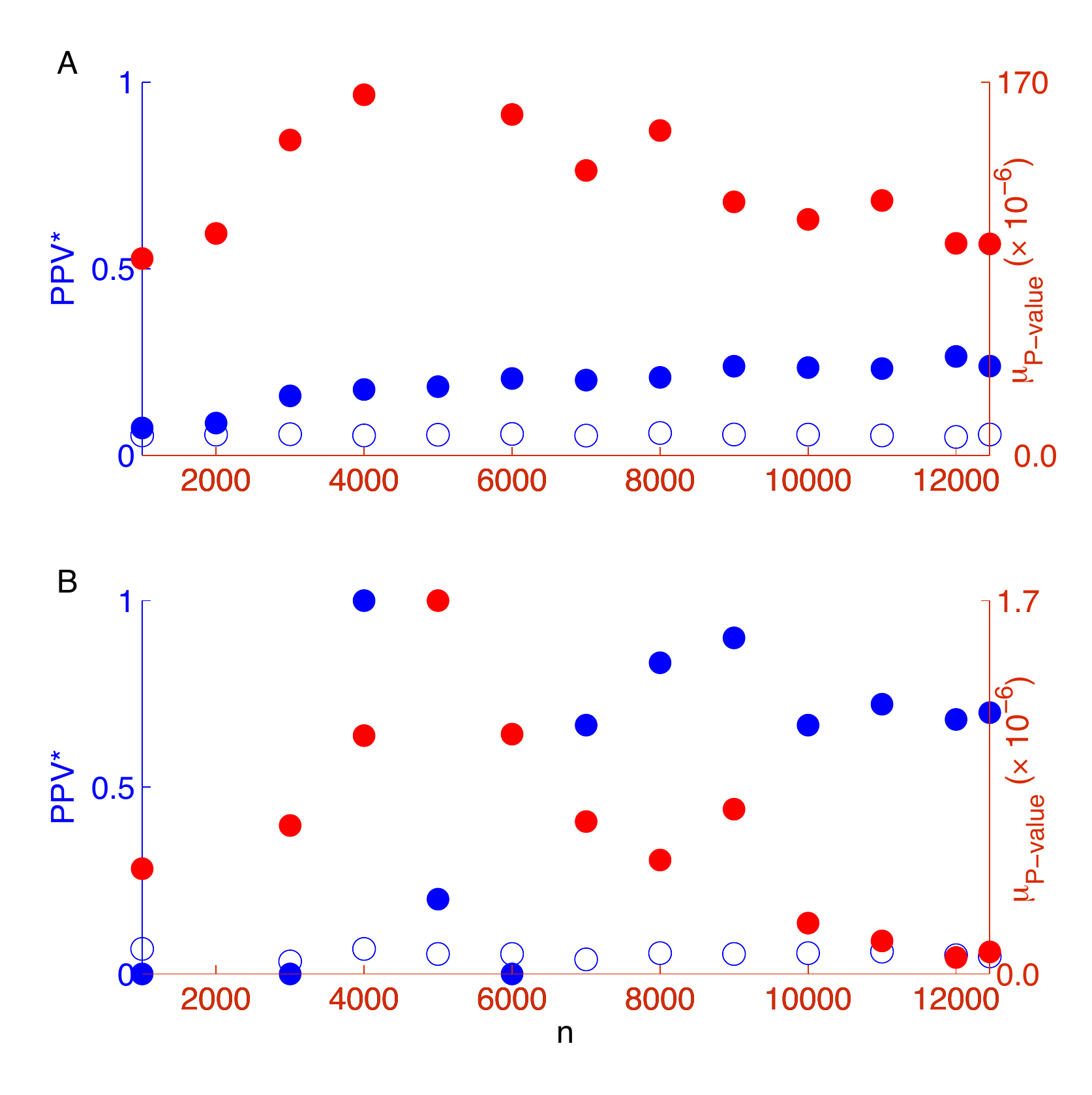}
    \caption{\textbf{Selection measures as a function of sample size in an analysis of real height data}. \textbf{(A)} The adjusted positive predictive value ($PPV^*$, blue solid dots) and median $P$-value ($\mu_\textrm{$P$-value}$, red) as a function of
    sample size using $\lambda$ based on $h^2=0.5$. Also shown is $PPV^*$ when the same number of SNPs are randomly selected rather than returned by the $L_1$ algorithm
    (blue unfilled dots). \textbf{(B)} As in (A) but setting $\lambda$ to a value appropriate for $h^2=0.01$.}\label{fig:height1}
\end{figure}

\begin{figure}[h!]
 \centering
\includegraphics[width=1.0\textwidth]{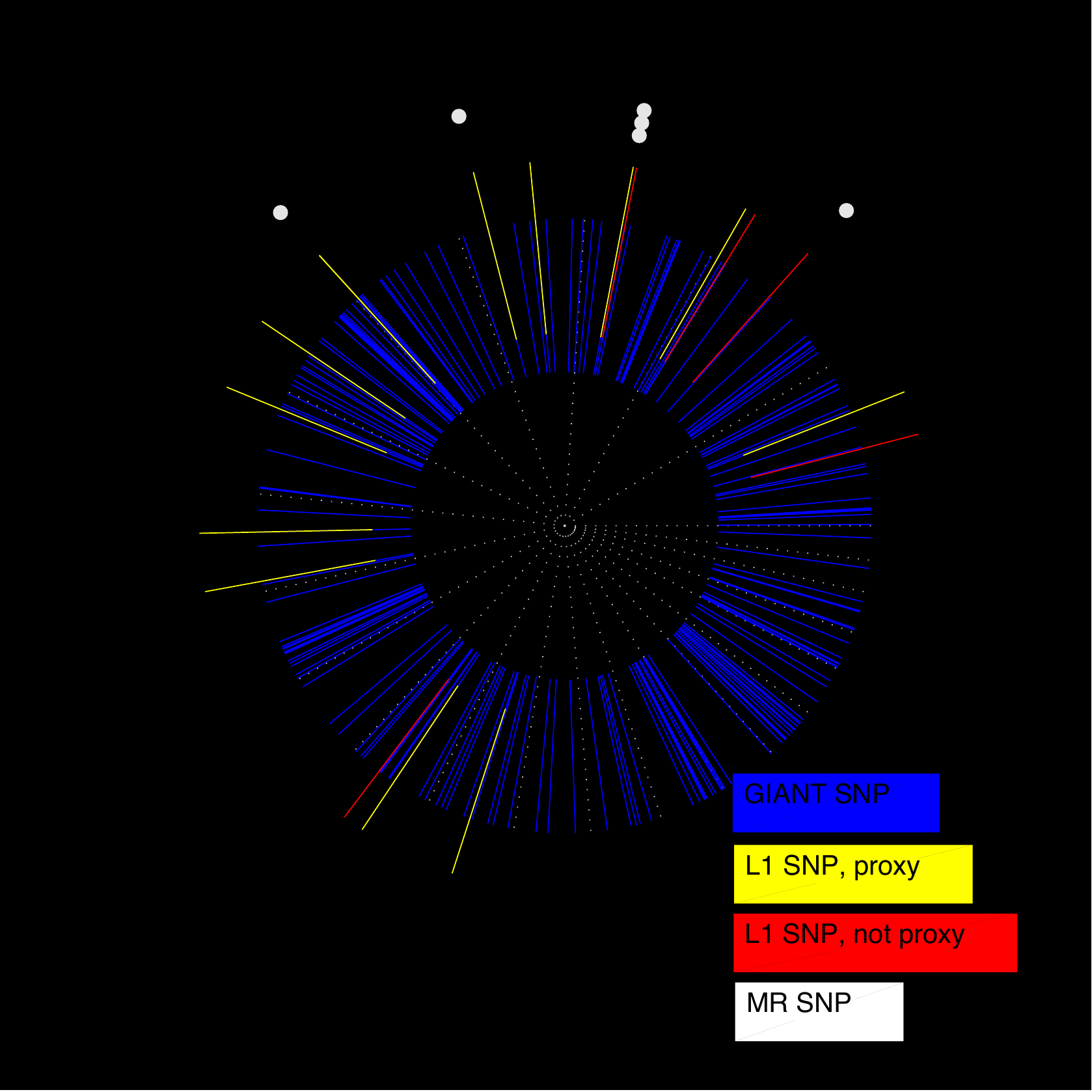}
    \caption{\textbf{Map of SNPs associated with height, as identified by the GIANT Consortium meta-analysis, $L_1$-penalized regression, and standard GWAS.} Base-pair distance is given by angle,
      and chromosome endpoints are demarcated by dotted lines. Starting from 3
      o'clock and going counterclockwise, the map sweeps through the
      chromosomes in numerical order. As a scale reference, the first
      sector represents chromosome 1 and is $\sim$ 250 million
      base-pairs. The blue segments correspond to a 1Mb window surrounding
      the height-associated
      SNPs discovered by GIANT. Note that some of these may overlap. The yellow segments represent $L_1$-selected SNPs that fell within 500 kb of a (blue) GIANT-identified nonzero; these met our
      criterion for being declared true positives. The red segments represent $L_1$-selected SNPs that did not fall
      within 500 kb of a GIANT-identified nonzero. Note that some yellow and red
      segments overlap given this figure's resolution. There
      are in total 20 yellow/red segments, representing $L_1$-selected SNPs found
      using all 12,454 subjects. The white dots represent the locations
      of SNPs selected by MR at a $P$-value threshold of $10^{-8}$.}\label{fig:height2}
\end{figure}

\begin{figure}[h!]
 \centering
\includegraphics[width=1.0\textwidth]{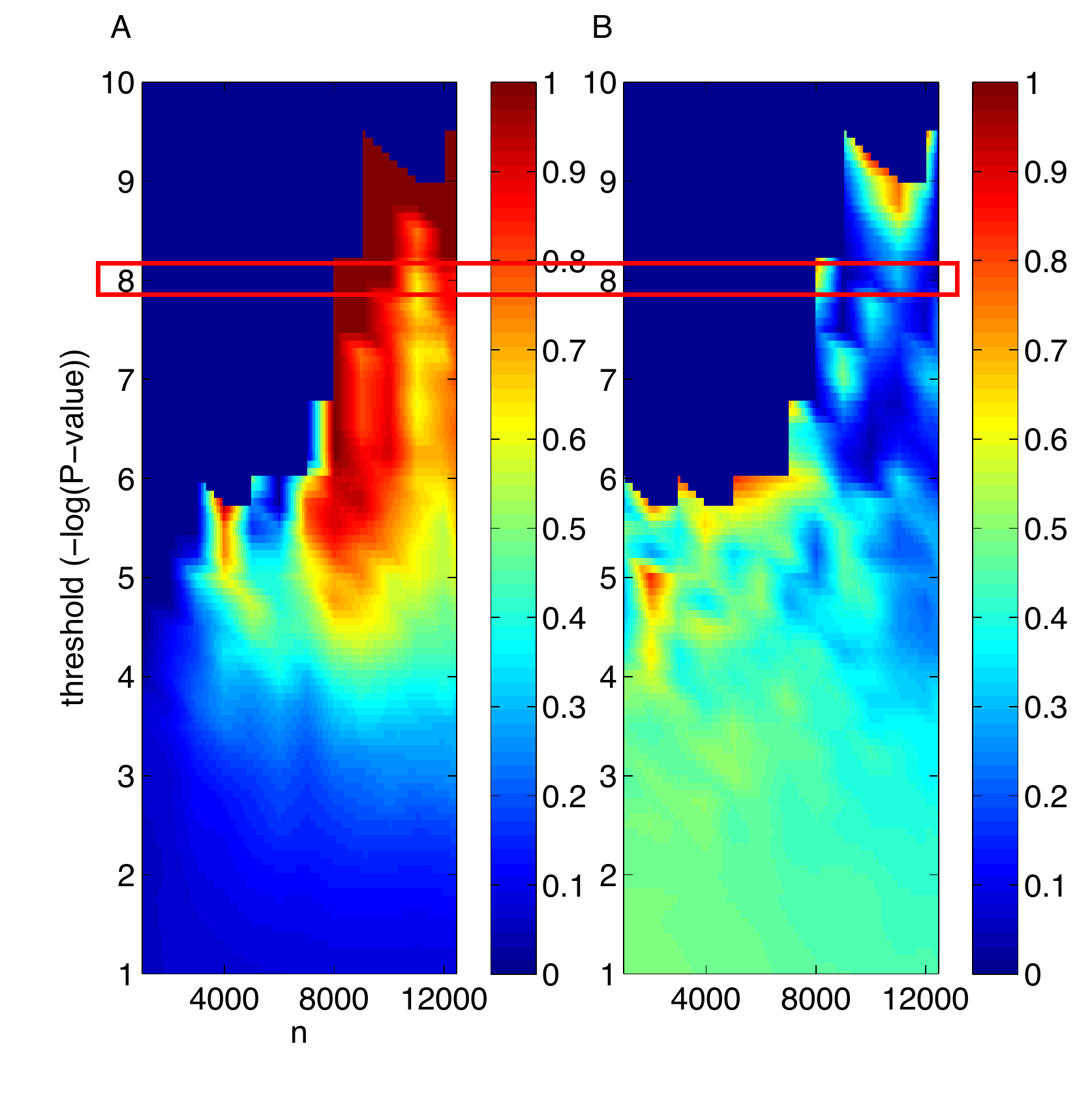}
    \caption{\textbf{Measures of recovery using marginal regression (standard GWAS) as a function of sample size.} All SNPs surviving the chosen $-\log_{10} P\textrm{-value}$ threshold were selected. The recovery measures, computed over the selected SNPs, were \textbf{(A)} the adjusted positive predictive value ($PPV^*$) and \textbf{(B)} the median $P$-value divided by the $P$-value cutoff. Highlighted in
    red is the cutoff we used for MR in Figure \ref{fig:height2}.}\label{fig:assoc}
\end{figure}

%
%
%
%

\end{backmatter}
\end{document}